\documentstyle[12pt,amsfonts]{article}
\newcommand{\svec}[1]{ \stackrel{\rightarrow}{#1} }

\newcommand{\ehat}{ \hat U_{\epsilon} }

\newcommand{\define}{ \stackrel{\triangle}{=} }

\def\be{\begin{equation}}
\def\ee{\end{equation}}
\def\ba{\begin{array}}
\def\ea{\end{array}}

\def\d4{{\rm d}^4}

\begin{document}
%---------------------------------------------------
\vskip -1.0cm
\title{\bf Classical Solution of Field Equation of Gravitational Gauge
            Field and Classical Tests of Gauge Theory of Gravity}
\author{ {Ning WU}\thanks{email address: wuning@mail.ihep.ac.cn}
{, Dahua ZHANG}
\\
\\
{\small Institute of High Energy Physics, P.O.Box 918-1,
Beijing 100039, P.R.China}
}
\maketitle
%\vskip 0.8in
%\noindent

%\vskip 0.8in
%\noindent

\begin{abstract}
A systematic method is developed to study classical motion of
a mass point in gravitational gauge field. First, the
formulation of gauge theory of gravity in arbitrary curvilinear
coordinates is given. Then in spherical coordinates system, a
spherical symmetric solution of the field equation of gravitational
gauge field is obtained, which is just the Schwarzschild solution.
In gauge theory of gravity, the equation of motion of
a classical mass point in gravitational gauge field is given
by Newton's second law of motion. A relativistic form of the
gravitational force on a mass point is deduced in this paper.
Based on the spherical symmetric solution of the field equation
and Newton's second law of motion, we can discuss classical
tests of gauge theory of gravity,
including the deflection of light by the sun,
the precession of the perihelia of the orbits of the inner planets
and the time delay of radar echoes passing the sun. It is found that
the theoretical predictions of these classical tests given by
gauge theory of gravity are completely the same as those given
by general relativity. From the study in this paper, an important
qualitative conclusion on the nature of  gravity is that
gravity can be treated as a kind of physical interactions
in flat Minkowski space-time, and the equation of motion
of mass point in gravitational field can be given by Newton's
second law of motion.
\end{abstract}

PACS Numbers:   04.80.Cc, 04.25.-g, 04.60.-m.  \\
Keywords: Classical tests of gauge theory of gravity,
        gauge theory of gravity,
        classical solution of field equation,
        Newton's second law of motion.   \\

%-------------------------------------------------------

\newpage

\Roman{section}

\section{Introduction}

It is known that Einstein's general relativity\cite{01,02} has passed
several classical tests, including the deflection of light
by the sun, the precession of the perihelia of the orbits of the
inner planets and the time delay of radar echoes passing
the sun\cite{03,04}. Basis of the calculation of these classical
tests in general relativity is Schwarzschild solution and geodesic
equation. Geodesic equation is given by the the shortest possible
path between two points, which is based on the concept of space-time
geometry. In Einstein's general theory of gravity, gravity is treated
as geometry of curved space-time.
\\

Quantum gauge theory of gravity\cite{05,06,07,08,09,10} is proposed
in the framework of quantum field theory. In gauge theory of gravity,
the field equation of gravitational gauge field is the same as
Einstein's field equation in general relativity, so two equations
have the same solutions, though mathematical expressions of the
two equations are completely different. Quantum
gauge theory of gravity is a perturbatively renormalizable
quantum theory, so based on it, quantum effects of
gravity\cite{11,12,13} and gravitational interactions of some basic
quantum fields \cite{14,15} can be explored. Unification of fundamental
interactions including gravity can be fulfilled in a simple and
beautiful way\cite{16,17,18}. If we use the mass
generation mechanism which is proposed
in literature \cite{19,20}, we can propose a new
theory on gravity which contains massive graviton and
the introduction of massive graviton does not affect
the strict local gravitational gauge symmetry of the
Lagrangian and does not affect the traditional long-range
gravitational force\cite{21}. The existence of massive graviton
will help us to understand the possible origin of dark matter.
\\

It is known that the transcendental foundations
of gauge theory of gravity is quite different from those of
general relativity. Basic concept in gauge theory of gravity
is that gravity is a kind of fundamental interactions in flat
Minkowski space-time, which is transmitted by gravitons. In other
words, in gauge theory of gravity, space-time is always flat, and
geodesic curve is always a straight line. So, in gauge theory
of gravity, geodesic equation can not be the equation of motion
of a mass point in gravitational field. In order to discussed
classical phenomenon of gravitational interactions, we need
first to set up the equation of motion of a mass point in
gravitational field, which is one of the central tasks of
this paper. Based on the spirit of gauge principle and completely
using physics language, Newton's second law of motion
is selected as an equation of motion for
a mass point. Combine this equation of motion
with classical solution of field equation, we can calculate
theoretical expectations of classical tests of gauge theory
of gravity. It is found that the quantitative values on classical
tests given  by gauge theory of gravity are completely the same
as those given by general relativity, though two theories use
different equations of motion in calculation and the basic
concepts on gravity in two theories are quite different.
 \\

\section{Basics of Gauge Theory of Gravity}
\setcounter{equation}{0}

For the sake of integrity, we give a simple
introduction to gauge theory of gravity and introduce
some notations which is used in this paper. Details on
quantum gauge theory of gravity can be found in
literatures \cite{05,06,07,08,09,10}.
In gauge theory of gravity, the most fundamental
quantity is gravitational gauge field $C_{\mu}(x)$,
which is the gauge potential corresponding to gravitational
gauge symmetry. Gauge field $C_{\mu}(x)$ is a vector in
the corresponding Lie algebra, which, for the sake
of convenience, will be called gravitational Lie algebra
in this paper. So $C_{\mu}(x)$ can be expanded as
\be \label{2.1}
C_{\mu}(x) = C_{\mu}^{\alpha} (x) \hat{P}_{\alpha},
~~~~~~(\mu, \alpha = 0,1,2,3)
\ee
where $C_{\mu}^{\alpha}(x)$ is the component field and
$\hat{P}_{\alpha} = -i \frac{\partial}{\partial x^{\alpha}}$
is the  generator of gravitational
gauge group, which satisfies
\be \label{2.2}
[ \hat{P}_{\alpha} ~~,~~ \hat{P}_{\beta} ]=0.
\ee
Unlike the ordinary $SU(N)$ group, the commutability of the
generators of the gravitational gauge group does not mean that the
gravitational gauge group is an Abelian group. In fact, the gravitational
gauge group is a non-Ablelian group\cite{05,06,07,08,09,10}.
The gravitational gauge covariant derivative is given by
\be \label{2.3}
D_{\mu} = \partial_{\mu} - i g C_{\mu} (x)
= G_{\mu}^{\alpha} \partial_{\alpha},
\ee
where $g$ is the gravitational coupling constant and matrix
$G$ is defined by
\be \label{2.4}
G = (G_{\mu}^{\alpha}) = ( \delta_{\mu}^{\alpha} - g C_{\mu}^{\alpha} ).
\ee
Matrix $G$ is an important quantity in gauge theory
of gravity. Its inverse matrix is denoted as $G^{-1}$
\be \label{2.5}
G^{-1} = \frac{1}{I - gC} = (G^{-1 \mu}_{\alpha}).
\ee
Using matrix $G$ and $G^{-1}$, we can define two important
composite operators
\be \label{2.6}
g^{\alpha \beta} = \eta^{\mu \nu}
G^{\alpha}_{\mu} G^{\beta}_{\nu},
\ee
\be \label{2.7}
g_{\alpha \beta} = \eta_{\mu \nu}
G_{\alpha}^{-1 \mu} G_{\beta}^{-1 \nu},
\ee
which are widely used in gauge theory of gravity. In gauge theory
of gravity, space-time is always flat and space-time metric
is always Minkowski metric, so $g^{\alpha\beta}$ and $g_{\alpha\beta}$
are no longer space-time metric. They are only two composite operators
which  consist of gravitational gauge field.
\\

The  field strength of gravitational gauge field is defined by
\be \label{2.8}
F_{\mu\nu} \define \frac{1}{-ig} \lbrack D_{\mu}~~,~~D_{\nu} \rbrack.
\ee
Its explicit expression is
\be \label{2.9}
F_{\mu\nu}(x) = \partial_{\mu} C_{\nu} (x)
-\partial_{\nu} C_{\mu} (x)
- i g C_{\mu} (x) C_{\nu}(x)
+ i g C_{\nu} (x) C_{\mu}(x).
\ee
$F_{\mu\nu}$ is also a vector in gravitational Lie algebra,
\be \label{2.10}
F_{\mu\nu} (x) = F_{\mu\nu}^{\alpha}(x) \cdot \hat{P}_{\alpha},
\ee
where
\be \label{2.11}
F_{\mu\nu}^{\alpha} = \partial_{\mu} C_{\nu}^{\alpha}
-\partial_{\nu} C_{\mu}^{\alpha}
-  g C_{\mu}^{\beta} \partial_{\beta} C_{\nu}^{\alpha}
+  g C_{\nu}^{\beta} \partial_{\beta} C_{\mu}^{\alpha}.
\ee
Using matrix $G$, its expression can be written in a
simpler form
\be \label{2.12}
F_{\mu\nu}^{\alpha} =
G_{\mu}^{\beta} \partial_{\beta} C_{\nu}^{\alpha}
-G_{\nu}^{\beta} \partial_{\beta} C_{\mu}^{\alpha}.
\ee
\\

In gauge theory of gravity, gravitational gauge field
$C_{\mu}^{\alpha}$ is a spin-2 tensor field, and the
lagrange for pure gravitational gauge field is selected
to be\cite{05,06}
\be
\ba{rcl}
{\cal L}_0
& = & - \frac{1}{16} \eta^{\mu \rho}
\eta^{\nu \sigma} g_{\alpha \beta}
F^{\alpha}_{\mu \nu} F^{\beta}_{\rho \sigma} \\
&&\\
&& - \frac{1}{8} \eta^{\mu \rho}
G^{-1 \nu}_{\beta} G^{-1 \sigma}_{\alpha}
F^{\alpha}_{\mu \nu} F^{\beta}_{\rho \sigma} \\
&&\\
&& + \frac{1}{4} \eta^{\mu \rho}
G^{-1 \nu}_{\alpha} G^{-1 \sigma}_{\beta}
F^{\alpha}_{\mu \nu} F^{\beta}_{\rho \sigma}.
\ea
\label{2.13}
\ee
The action is defined by
\be
S = \int {\rm d}^4 x \sqrt{- {\rm det} (g_{\alpha \beta}) }
\cdot {\cal L}_0.
\label{2.14}
\ee
It can be proved that the above action $S$ is invariant  under
gravitational gauge transformation, therefore the system has
gravitational gauge symmetry\cite{05,06}.
\\

The Euler-Lagrange equation for gravitational gauge field is
\be
\partial_{\mu} \frac{\partial {\cal L }_0}
{\partial \partial_{\mu} C_{\nu}^{\alpha}}
= \frac{\partial {\cal L}_0}{\partial C_{\nu}^{\alpha}}
+g G_{\alpha}^{-1 \nu} {\cal L}_0
- g G_{\gamma}^{-1 \lambda}
(  \partial_{\mu} C_{\lambda}^{\gamma})
\frac{\partial {\cal L}_0}{\partial \partial_{\mu} C_{\nu}^{\alpha}},
\label{2.15}
\ee
which gives  the following field equation of gravitational gauge field
\be
\ba{rl}
\partial_{\mu} ( & \frac{1}{4} \eta^{\mu \rho}
\eta^{\nu \sigma} g_{\alpha \beta}
F_{\rho \sigma}^{\beta}
- \frac{1}{4} \eta^{\nu \rho} F^{\mu}_{\rho\alpha}
+ \frac{1}{4} \eta^{\mu \rho} F^{\nu}_{\rho\alpha} \\
&\\
& - \frac{1}{2} \eta^{\mu\rho} \delta^{\nu}_{\alpha}
F^{\beta}_{\rho\beta}
+ \frac{1}{2} \eta^{\nu\rho} \delta^{\mu}_{\alpha}
F^{\beta}_{\rho\beta})  = - g T_{g \alpha}^{\nu},
\label{2.16}
\ea
\ee
where
\be  \label{2.17}
\ba{rcl}
T_{g \alpha}^{\nu} & = &
 \frac{1}{4} \eta^{\nu\rho} \eta^{\lambda\sigma}
g_{\beta\gamma} F^{\beta}_{\rho\sigma}
(\partial_{\alpha} C_{\lambda}^{\gamma}) \\
&& + \frac{1}{4} \eta^{\nu\rho} G^{-1 \lambda}_{\beta}
G^{-1 \sigma}_{\gamma} F^{\beta}_{\rho\sigma}
(\partial_{\alpha} C_{\lambda}^{\gamma}) \\
&& - \frac{1}{2} \eta^{\nu\rho} G^{-1 \lambda}_{\gamma}
G^{-1 \sigma}_{\beta} F^{\beta}_{\rho\sigma}
(\partial_{\alpha} C_{\lambda}^{\gamma}) \\
&& - \frac{1}{4} \eta^{\lambda\rho} G^{-1 \nu}_{\beta}
G^{-1 \sigma}_{\gamma} F^{\beta}_{\rho\sigma}
(\partial_{\alpha} C_{\lambda}^{\gamma}) \\
&& + \frac{1}{2} \eta^{\lambda\rho} G^{-1 \nu}_{\gamma}
G^{-1 \sigma}_{\beta} F^{\beta}_{\rho\sigma}
(\partial_{\alpha} C_{\lambda}^{\gamma}) \\
&& + \frac{1}{4} \eta^{\lambda\rho} \eta^{\nu\sigma}
g_{\alpha\beta}
G^{-1 \kappa}_{\gamma} F^{\beta}_{\rho\sigma}
(D_{\lambda} C_{\kappa}^{\gamma}) \\
&& - \frac{1}{4} \eta^{\nu\rho} G^{-1 \sigma}_{\alpha}
G^{-1 \kappa}_{\gamma} F^{\mu}_{\rho\sigma}
(\partial_{\mu} C_{\kappa}^{\gamma}) \\
&& + \frac{1}{2} \eta^{\nu\rho} G^{-1 \sigma}_{\beta}
G^{-1 \kappa}_{\gamma} F^{\beta}_{\rho\sigma}
(\partial_{\alpha} C_{\kappa}^{\gamma}) \\
&& + \frac{1}{4} \eta^{\lambda\rho} G^{-1 \nu}_{\beta}
G^{-1 \sigma}_{\alpha}
G^{-1 \kappa}_{\gamma} F^{\beta}_{\rho\sigma}
(D_{\lambda} C_{\kappa}^{\gamma}) \\
&& - \frac{1}{2} \eta^{\lambda\rho} G^{-1 \nu}_{\alpha}
G^{-1 \sigma}_{\beta}
G^{-1 \kappa}_{\gamma} F^{\beta}_{\rho\sigma}
(D_{\lambda} C_{\kappa}^{\gamma}) \\
&& - \frac{1}{4} \eta^{\lambda\rho} \eta^{\nu\sigma}
\partial_{\mu} (g_{\alpha \beta} C^{\mu}_{\lambda}
 F^{\beta}_{\rho\sigma}) \\
&& - \frac{1}{4} \eta^{\nu\rho}
\partial_{\beta} (C^{\sigma}_{\lambda} G^{-1 \lambda}_{\alpha}
 F^{\beta}_{\rho\sigma}) \\
&&  + \frac{1}{2} \eta^{\nu\rho}
\partial_{\alpha} (C^{\sigma}_{\lambda} G^{-1 \lambda}_{\beta}
 F^{\beta}_{\rho\sigma}) \\
&& + \frac{1}{4g} \eta^{\lambda\rho}
\partial_{\mu} \lbrack ( G^{-1 \nu}_{\beta} G^{-1 \sigma}_{\alpha}
G^{\mu}_{\lambda} - \delta^{\nu}_{\beta} \delta^{\sigma}_{\alpha}
\delta^{\mu}_{\lambda}) F^{\beta}_{\rho\sigma}
\rbrack  \\
&& - \frac{1}{2 g} \eta^{\lambda\rho}
\partial_{\mu} \lbrack ( G^{-1 \nu}_{\alpha} G^{-1 \sigma}_{\beta}
G^{\mu}_{\lambda} - \delta^{\nu}_{\alpha} \delta^{\sigma}_{\beta}
\delta^{\mu}_{\lambda}) F^{\beta}_{\rho\sigma}
\rbrack  \\
&& - \frac{1}{4} \eta^{\kappa \rho} G^{-1 \nu}_{\beta}
G^{-1 \lambda}_{\alpha} G^{-1 \sigma}_{\gamma}
F^{\beta}_{\rho\sigma} F^{\gamma}_{\kappa\lambda}  \\
&& + \frac{1}{2} \eta^{\kappa \rho} G^{-1 \nu}_{\gamma}
G^{-1 \lambda}_{\alpha} G^{-1 \sigma}_{\beta}
F^{\beta}_{\rho\sigma} F^{\gamma}_{\kappa\lambda}  \\
&& - \frac{1}{8} \eta^{\mu \rho} \eta^{\lambda \sigma}
g_{\alpha \gamma} G^{-1 \nu}_{\beta}
F^{\beta}_{\rho\sigma} F^{\gamma}_{\mu\lambda}  \\
&& - \frac{1}{16} \eta^{\mu \rho} \eta^{\lambda \sigma}
g_{\beta \gamma} G^{-1 \nu}_{\alpha}
F^{\beta}_{\rho\sigma} F^{\gamma}_{ \mu\lambda}  \\
&& - \frac{1}{8} \eta^{\mu \rho} G^{-1 \nu}_{\alpha}
G^{-1 \lambda}_{\beta} G^{-1 \sigma}_{\gamma}
F^{\beta}_{\rho\sigma} F^{\gamma}_{\mu\lambda}  \\
&& + \frac{1}{4} \eta^{\mu \rho} G^{-1 \nu}_{\alpha}
G^{-1 \lambda}_{\gamma} G^{-1 \sigma}_{\beta}
F^{\beta}_{\rho\sigma} F^{\gamma}_{\mu\lambda}.
\ea
\ee
$T_{g \alpha}^{\nu}$ is the gravitational energy-momentum
tensor, which is the source of gravitational field. It can be proved
that this field equation is the same as the Einstein's field
equation\cite{05,06}.
\\

Gauge theory of gravity is formulated in physics picture of gravity,
where gravity is treated as a kind of fundamental interactions and
space-time is always flat. But for classical problems, we can also
set up a geometry picture of gravity, where gravity is equivalently
treated as geometry of space-time\cite{22}. In this equivalent
space-time geometry, $g_{\alpha\beta}$ and $g^{\alpha\beta}$ are
equivalent metric of the curved space-time. From this equivalent metric,
we can calculate affine connection
\be
\Gamma^{\prime\gamma}_{\alpha \beta}
= \frac{1}{2} g^{\gamma \delta}
\left( \frac{\partial g_{\alpha \delta}}{\partial x^{\beta}}
+ \frac{\partial g_{\beta \delta}}{\partial x^{\alpha}}
-\frac{\partial g_{\alpha \beta}}{\partial x^{\delta}} \right),
\label{2.18}
\ee
and curvature tensor
\be
R^{\delta}_{\alpha \beta \gamma}
\define \partial_{\gamma} \Gamma^{\prime\delta}_{\alpha \beta}
-\partial_{\beta} \Gamma^{\prime\delta}_{\alpha \gamma}
+\Gamma^{\prime\eta}_{\alpha \beta} \Gamma^{\prime\delta}_{\gamma \eta}
- \Gamma^{\prime\eta}_{\alpha \gamma} \Gamma^{\prime\delta}_{\beta \eta}.
\label{2.19}
\ee
Recci tensor $R_{\alpha \gamma}$ is defined by
\be
R_{\alpha \gamma} \define
R^{\beta}_{\alpha \beta \gamma}.
\label{2.20}
\ee
Operator $R_{\alpha \gamma}$ can also be calculated from the
following relation
\be
\ba{rcl}
R_{\alpha \gamma}
& = & \frac{1}{2} g^{\beta \delta}
(\partial_{\beta} \partial_{\delta} g_{\alpha \gamma}
- \partial_{\alpha} \partial_{\beta} g_{\delta \gamma}
-\partial_{\delta} \partial_{\gamma} g_{\alpha \beta}
+\partial_{\alpha} \partial_{\gamma} g_{\beta \delta} ) \\
&&\\
&& + g^{\beta \delta} g_{\alpha_1 \beta_1}
(\Gamma^{\prime\alpha_1}_{\alpha \gamma}
\Gamma^{\prime\beta_1}_{\beta \delta}
- \Gamma^{\prime\alpha_1}_{\alpha \beta}
\Gamma^{\prime\beta_1}_{\delta \gamma}).
\ea
\label{2.21}
\ee
After rather lengthy and  complicated calculations, we
can proved that the operator  $R_{\alpha \beta}$ can be
explicitly expressed as
\be
\begin{array}{rcl}
R_{\alpha \beta} &=&
- \frac{g^2}{4} \eta^{\mu\rho} \eta^{\nu \sigma}
g_{\alpha\alpha_1} g_{\beta \beta_1}
F_{\mu\nu}^{\alpha_1} F_{\rho\sigma}^{\beta_1}\\
&&\\
&& + \frac{g^2}{2} \eta^{\nu\sigma} g_{\alpha_1 \beta_1}
G^{-1 \rho}_{\alpha} G^{-1 \mu}_{\beta}
F_{\mu\nu}^{\alpha_1} F_{\rho\sigma}^{\beta_1}  \\
&&\\
&& + \frac{g^2}{2}
G^{-1 \mu}_{\beta}  G^{-1 \rho }_{\alpha}
G^{-1 \nu}_{\beta_1} G^{-1 \sigma}_{\alpha_1}
F_{\mu\nu}^{\alpha_1} F_{\rho\sigma}^{\beta_1}\\
&&\\
&& + \frac{g^2}{2} \eta^{\nu\sigma}
(g_{\alpha_1 \alpha} G^{-1 \mu}_{\beta} +
g_{\alpha_1 \beta} G^{-1 \mu}_{\alpha} )
G^{-1 \rho}_{\beta_1}
F_{\mu\nu}^{\alpha_1} F_{\rho\sigma}^{\beta_1} \\
&&\\
&& + \frac{g}{2}
(G^{-1 \mu}_{\beta} \partial_{\alpha} +
G^{-1 \mu}_{\alpha} \partial_{\beta})
(G^{-1 \nu}_{\gamma} F_{\mu\nu}^{\gamma}) \\
&&\\
&& +\frac{g}{2} \eta_{\mu_1 \nu_1} g^{\delta \beta_1}
(G^{-1 \mu_1}_{\alpha} G^{-1 \nu}_{\beta}+
G^{-1 \mu_1}_{\beta} G^{-1 \nu}_{\alpha})
G^{-1 \mu}_{\beta_1} \partial_{\delta}
(G^{-1 \nu_1}_{\gamma} F_{\mu\nu}^{\gamma}).
\end{array}
\label{2.22}
\ee
The scalar curvature $R$ is defined by
\be
R \define g^{\alpha \beta} R_{\alpha \beta}.
\label{2.23}
\ee
The explicit expression for the operator $R$ is
\be
R = R_0 + \frac{2g}{J(C)} \partial_{\beta}
\left (
 J(C) g^{\alpha \beta} G^{-1 \mu}_{\alpha}
G^{-1 \nu}_{\gamma} F^{\gamma}_{\mu\nu}
\right ),
\label{2.24}
\ee
where
\be \label{2.25}
R_0 = - 16 \pi G  {\cal L}_0,
\ee
with $G$ the Newtonian gravitation constant. Its relation
to gravitation coupling constant $g$ is given by the
following formula
\be \label{2.26}
G= \frac{g^2}{4 \pi}.
\ee
Because total derivative term in the Lagrangian has no
contribution to Euler-Lagrange equation of motion, selecting
${\cal L}_0$ as Lagrangian is equivalent to selecting
$- \frac{1}{16 \pi G} R$ as Lagrangian. In general relativity,
$- \frac{1}{16 \pi G} R$ is selected as Lagrangian of gravitational
field, the field equation given by this selection is the Einstein's
field equation
\be
R_{\alpha \beta} - \frac{1}{2} g_{\alpha \beta} R
+ 8 \pi G T_{\alpha \beta} =0,
\label{2.27}
\ee
where $T_{\alpha \beta}$ is the energy-momentum tensor of
matter field.   \\

Now, from the
same action, the least action principle gives  two equations
(\ref{2.16}) and (\ref{2.27}). They are different in forms,
but they are essentially the same, for one action can only give
only one field equation. In deed, we can strictly prove that
these two field equations are essentially the same. If define
\be
W^{\nu}_{\alpha} =
- \partial_{\mu} \frac{\partial {\cal L }_0}
{\partial \partial_{\mu} C_{\nu}^{\alpha}}
+\frac{\partial {\cal L}_0}{\partial C_{\nu}^{\alpha}}
+g G_{\alpha}^{-1 \nu} {\cal L}_0
- g G_{\gamma}^{-1 \lambda}
(  \partial_{\mu} C_{\lambda}^{\gamma})
\frac{\partial {\cal L}_0}{\partial \partial_{\mu} C_{\nu}^{\alpha}},
\label{2.28}
\ee
then field equation (\ref{2.16}) can be simply expressed as
\be
W^{\nu}_{\alpha} =0.
\label{2.29}
\ee
It can be strictly proved that
\be
W^{\nu}_{\alpha} \eta_{\mu\nu} G^{-1 \mu}_{\beta}
= \frac{1}{2g} (R_{\alpha\beta}
-\frac{1}{2} g_{\alpha\beta} R).
\label{2.30}
\ee
Therefor, field equation (\ref{2.29}) just gives  the Einstein's
field equation. So, in quantum gauge general relativity,
the field equation of gravitational gauge field is just the Einstein's
field equation.
\\

\section{Formulation of Gauge Theory of Gravity in Arbitrary
            Curvilinear Coordinate System}
\setcounter{equation}{0}

Gauge theory of gravity is proposed in physics picture of
gravity, where space-time is always flat. All above discussions
are performed in Descartes coordinate system and the above
expressions of mathematical formula are only valid in
Descartes coordinate system.
In order to solve spherical symmetric classical solution of
field equation (\ref{2.16}), we need to formulate gauge theory
of gravity in arbitrary curvilinear coordinate system.
\\

Denote space-time coordinates in Descartes coordinate system
as $x^{\mu}$. Making the following coordinates transformations
\be \label{3.1}
x \to y=y(x),
\ee
where $y^{\mu}$ are arbitrary curvilinear coordinates. Obviously, we have
\be \label{3.2}
{\rm d} y^{\alpha_1}
= \frac{\partial y^{\alpha_1}}{\partial x^{\alpha}}~
{\rm d} x^{\alpha},
\ee
\be \label{3.3}
\frac{\partial }{\partial y^{\alpha_1}}
= \frac{\partial x^{\alpha}}{\partial y^{\alpha_1}}~
\frac{\partial }{\partial x^{\alpha}}.
\ee
In this chapter, indexes $\alpha$, $\beta$, $\gamma$, $\mu$, $\nu$
$\lambda$, $\cdots$ are used to denote indexes of Descartes coordinate
system, while indexes $\alpha_1$, $\beta_1$, $\gamma_1$, $\mu_1$, $\nu_1$
$\lambda_1$, $\cdots$ are used to denote indexes of curvilinear
coordinate system. This convention is only valid in this chapter,
it is no longer valid in other chapters of this paper. \\

Suppose that $A^{\alpha}$ and $B_{\beta}$ are two arbitrary
vectors, under coordinate transformation (\ref{3.1}),
they transforms as
\be \label{3.4}
A^{\alpha} \to A^{\prime \alpha_1}
= \frac{\partial y^{\alpha_1}}{\partial x^{\alpha}}~
A^{\alpha},
\ee
\be \label{3.5}
B_{\alpha} \to B'_{\alpha_1}
= \frac{\partial x^{\alpha}}{\partial y^{\alpha_1}}~
B_{\alpha}.
\ee
Space-time metrics $\eta^{\mu\nu}$ and $\eta_{\mu\nu}$ are second
order tensors, so they transform as
\be \label{3.6}
\eta^{\mu\nu} \to \eta^{\prime \mu_1 \nu_1}
= \frac{\partial y^{\mu_1}}{\partial x^{\mu}}
\frac{\partial y^{\nu_1}}{\partial x^{\nu}}
\eta^{\mu \nu},
\ee
\be \label{3.7}
\eta_{\mu \nu} \to \eta'_{\mu_1 \nu_1}
= \frac{\partial x^{\mu}}{\partial y^{\mu_1}}
\frac{\partial x^{\nu}}{\partial y^{\nu_1}}
\eta_{\mu \nu},
\ee
where $\eta^{\prime \mu_1 \nu_1}$ and $\eta'_{\mu_1 \nu_1}$ are space-time
metric in curvilinear coordinate system. The affine connection
in curvilinear coordinate system is defined by
\be
\Gamma^{\lambda_1}_{\mu_1 \nu_1}
= \frac{1}{2} \eta^{\prime \lambda_1 \sigma_1}
\left( \frac{\partial \eta'_{\mu_1 \sigma_1}}{\partial y^{\nu_1}}
+ \frac{\partial \eta'_{\nu_1 \sigma_1}}{\partial y^{\mu_1}}
-\frac{\partial \eta'_{\mu_1 \nu_1}}{\partial y^{\sigma_1}} \right).
\label{3.8}
\ee
It can be expressed in other forms
\be
\ba{rcl}
\Gamma^{\lambda_1}_{\mu_1 \nu_1} &= &
 \frac{\partial y^{\lambda_1}}{\partial x^{\nu}}
 \frac{\partial}{\partial y^{\mu_1}}
 \frac{\partial x^{\nu}}{\partial y^{\nu_1}} \\
 &&\\
&= & - \frac{\partial x^{\nu}}{\partial y^{\nu_1}}
\frac{\partial x^{\mu}}{\partial y^{\mu_1}}
 \frac{\partial^2 y^{\lambda_1}}{\partial x^{\mu} \partial x^{\nu}}.
\label{3.9}
\ea
\ee
\\

Using above relations, we can prove that
\be  \label{3.10}
\frac{ \partial A^{\mu} }{\partial x^{\alpha}} =
\frac{\partial x^{\mu}}{\partial y^{\mu_1}}
\frac{\partial y^{\alpha_1}}{\partial x^{\alpha}}
(\nabla_{\alpha_1} A^{\prime \mu_1}),
\ee
\be  \label{3.11}
\frac{ \partial B_{\mu} }{\partial x^{\alpha}} =
\frac{\partial y^{\mu_1}}{\partial x^{\mu}}
\frac{\partial y^{\alpha_1}}{\partial x^{\alpha}}
(\nabla_{\alpha_1} B'_{\mu_1}),
\ee
\be  \label{3.12}
\frac{{\rm d} }{{\rm d } \tau}  A^{\mu}  =
\frac{\partial x^{\mu}}{\partial y^{\mu_1}}
\frac{D}{ D \tau}  A^{\prime \mu_1},
\ee
\be  \label{3.13}
\frac{{\rm d} }{{\rm d } \tau}  B_{\mu}  =
\frac{\partial y^{\mu_1}}{\partial x^{\mu}}
\frac{D}{ D \tau}  B'_{\mu_1},
\ee
where $\nabla_{\alpha_1}$ and $\frac{D}{D \tau}$ are the
covariant derivatives in curvilinear coordinate system, which
are defined by
\be  \label{3.14}
\nabla_{\alpha_1} A^{\prime \mu_1} =
\frac{\partial A^{\prime \mu_1}}{\partial y^{\alpha_1}}
+ \Gamma^{\mu_1}_{\alpha_1 \nu_1} A^{\prime \nu_1},
\ee
\be  \label{3.15}
\nabla_{\alpha_1} B'_{\mu_1} =
\frac{\partial B'_{\mu_1}}{\partial y^{\alpha_1}}
- \Gamma^{\nu_1}_{\alpha_1 \mu_1} B'_{\nu_1},
\ee
\be  \label{3.16}
\frac{D}{D \tau } A^{\prime \mu_1}  =
\frac{\rm d}{{\rm d} \tau} A^{\prime \mu_1}
+ \Gamma^{\mu_1}_{\alpha_1 \nu_1}
\frac{{\rm d} y^{\alpha_1}}{{\rm d} \tau} A^{\prime \nu_1},
\ee
\be  \label{3.17}
\frac{D}{D \tau } B'_{\mu_1}  =
\frac{\rm d}{{\rm d} \tau} B'_{\mu_1}
- \Gamma^{\nu_1}_{\alpha_1 \mu_1}
\frac{{\rm d} y^{\alpha_1}}{{\rm d} \tau} B'_{\nu_1}.
\ee
Therefore, under curvilinear transformations (\ref{3.1}),
we have
\be  \label{3.18}
\frac{\partial A^{\mu}}{\partial x^{\alpha}} \to
 \nabla_{\alpha_1} A^{\prime \mu_1} =
\frac{\partial y^{\mu_1}}{\partial x^{\mu}}
\frac{\partial x^{\alpha}}{\partial y^{\alpha_1}}
\left ( \frac{\partial A^{\mu}}{\partial x^{\alpha}} \right ),
\ee
\be  \label{3.19}
\frac{\partial B_{\mu}}{\partial x^{\alpha}} \to
 \nabla_{\alpha_1} B'_{\mu_1} =
\frac{\partial x^{\mu}}{\partial y^{\mu_1}}
\frac{\partial x^{\alpha}}{\partial y^{\alpha_1}}
\left ( \frac{\partial B_{\mu}}{\partial x^{\alpha}} \right ),
\ee
\be  \label{3.20}
\frac{\rm d}{{\rm d} \tau} A^{\mu} \to
 \frac{D}{D \tau} A^{\prime \mu_1} =
\frac{\partial y^{\mu_1}}{\partial x^{\mu}}
\left ( \frac{\rm d}{{\rm d} \tau}  A^{\mu} \right ),
\ee
\be  \label{3.21}
\frac{\rm d}{{\rm d} \tau} B_{\mu} \to
 \frac{D}{D \tau} B'_{\mu_1} =
\frac{\partial x^{\mu }}{\partial y^{\mu_1}}
\left ( \frac{\rm d}{{\rm d} \tau}  B_{\mu} \right ).
\ee
\\

Gravitational gauge field $C_{\mu}^{\alpha}$ is a second order
tensor under curvilinear coordinate transformations, therefore
its transformation is
\be  \label{3.22}
C_{\mu}^{\alpha } \to C_{\mu_1}^{\prime \alpha_1}
 = \frac{\partial x^{\mu }}{\partial y^{\mu_1}}
\frac{\partial y^{\alpha_1 }}{\partial x^{\alpha}}
C_{\mu}^{\alpha}.
\ee
Then, from equations (\ref{2.4}) and (\ref{2.5}), we can prove
that
\be  \label{3.23}
G_{\mu}^{\alpha } \to G_{\mu_1}^{\prime \alpha_1}
 = \frac{\partial x^{\mu }}{\partial y^{\mu_1}}
\frac{\partial y^{\alpha_1 }}{\partial x^{\alpha}}
G_{\mu}^{\alpha},
\ee
\be  \label{3.24}
G^{-1 \mu}_{\alpha } \to G^{\prime -1 \mu_1}_{ \alpha_1}
 = \frac{\partial y^{\mu_1 }}{\partial x^{\mu}}
\frac{\partial x^{\alpha  }}{\partial y^{\alpha_1}}
G^{-1 \mu}_{\alpha}.
\ee
Applying the above two relations and equations
(\ref{2.6}) and (\ref{2.7}), we find that
\be \label{3.25}
g^{\alpha\beta} \to g^{\prime \alpha_1 \beta_1}
= \frac{\partial y^{\alpha_1}}{\partial x^{\alpha}}
\frac{\partial y^{\beta_1}}{\partial x^{\beta}}
g^{\alpha \beta},
\ee
\be \label{3.26}
g_{\alpha \beta} \to g'_{\alpha_1 \beta_1}
= \frac{\partial x^{\alpha}}{\partial y^{\alpha_1}}
\frac{\partial x^{\beta}}{\partial y^{\beta_1}}
g_{\alpha \beta}.
\ee
\\

Using equations (\ref{3.18}), (\ref{3.19}) and (\ref{3.22}),
we can prove that
\be \label{3.27}
\frac{\partial}{\partial x^{\beta}} C_{\mu}^{\alpha}
\to \nabla_{\beta_1} C_{\mu_1}^{\prime \alpha_1} =
\frac{\partial y^{\alpha_1}}{\partial x^{\alpha}}
\frac{\partial x^{\beta}}{\partial y^{\beta_1}}
\frac{\partial x^{\mu}}{\partial y^{\mu_1}}
\left ( \frac{\partial}{\partial x^{\beta}} C_{\mu}^{\alpha}
\right ),
\ee
\be \label{3.28}
F_{\mu \nu}^{\alpha}
\to F_{\mu_1 \nu_1}^{\prime \alpha_1} =
\frac{\partial y^{\alpha_1}}{\partial x^{\alpha}}
\frac{\partial x^{\mu}}{\partial y^{\mu_1}}
\frac{\partial x^{\nu}}{\partial y^{\nu_1}}
F_{\mu \nu}^{\alpha},
\ee
where
\be \label{3.29}
\nabla_{\beta_1} C_{\mu_1}^{\prime \alpha_1} =
\frac{\partial}{\partial y_{\beta_1}} C_{\mu_1}^{\prime \alpha_1}
+ \Gamma^{\alpha_1}_{\beta_1 \gamma_1} C_{\mu_1}^{\prime \gamma_1}
- \Gamma^{\nu_1}_{\beta_1 \mu_1} C_{\nu_1}^{\prime \alpha_1},
\ee
\be \label{3.30}
\ba{rcl}
F_{\mu_1 \nu_1}^{\prime \alpha_1} & = &
G_{\mu_1}^{\prime \beta_1} \nabla_{\beta_1} C_{\nu_1}^{\prime \alpha_1}
-G_{\nu_1}^{\prime \beta_1} \nabla_{\beta_1} C_{\mu_1}^{\prime \alpha_1} \\
&=& G_{\mu_1}^{\prime \beta_1} \left \lbrack
\frac{\partial}{\partial y_{\beta_1}} C_{\nu_1}^{\prime \alpha_1}
+ \Gamma^{\alpha_1}_{\beta_1 \gamma_1} C_{\nu_1}^{\prime \gamma_1}
- \Gamma^{\lambda_1}_{\beta_1 \nu_1} C_{\lambda_1}^{\prime \alpha_1} \right \rbrack \\
&& - G_{\nu_1}^{\prime \beta_1} \left \lbrack
\frac{\partial}{\partial y_{\beta_1}} C_{\mu_1}^{\prime \alpha_1}
+ \Gamma^{\alpha_1}_{\beta_1 \gamma_1} C_{\mu_1}^{\prime \gamma_1}
- \Gamma^{\lambda_1}_{\beta_1 \mu_1} C_{\lambda_1}^{\prime \alpha_1} \right \rbrack.
\ea
\ee
\\

In conclusion, under curvilinear coordinate
transformations, all physical quantities transforms covariantly.
Physical equations will not change their forms under this transformation
if we replace ordinary derivatives $\partial_{\alpha}$ and
$\frac{\rm d}{{\rm d} \tau}$ with covariant derivatives
$\nabla_{\alpha}$ and $\frac{D}{D \tau}$. Using this way,
we can formulate gauge theory of gravity in arbitrary
curvilinear coordinate system. \\

\section{Classical Spherical Symmetric Solution of Field Equation
            of Gravitational Gauge Field }
\setcounter{equation}{0}

In the last chapter, formulation of gauge theory of gravity
in arbitrary curvilinear coordinate system is discussed. Based
on the results in the last chapter, we can obtain the following
field equation of gravitational gauge field in arbitrary
curvilinear coordinate system from equation (\ref{2.16})
\be
\ba{rl}
\nabla_{\mu} ( & \frac{1}{4} \eta^{\mu \rho}
\eta^{\nu \sigma} g_{\alpha \beta}
F_{\rho \sigma}^{\beta}
- \frac{1}{4} \eta^{\nu \rho} F^{\mu}_{\rho\alpha}
+ \frac{1}{4} \eta^{\mu \rho} F^{\nu}_{\rho\alpha} \\
&\\
& - \frac{1}{2} \eta^{\mu\rho} \delta^{\nu}_{\alpha}
F^{\beta}_{\rho\beta}
+ \frac{1}{2} \eta^{\nu\rho} \delta^{\mu}_{\alpha}
F^{\beta}_{\rho\beta})  = - g T_{g \alpha}^{\nu},
\label{4.1}
\ea
\ee
where gravitational energy-momentum tensor $T_{g \alpha}^{\nu}$
is given by
\be  \label{4.2}
\ba{rcl}
T_{g \alpha}^{\nu} & = &
 \frac{1}{4} \eta^{\nu\rho} \eta^{\lambda\sigma}
g_{\beta\gamma} F^{\beta}_{\rho\sigma}
(\nabla_{\alpha} C_{\lambda}^{\gamma}) \\
&& + \frac{1}{4} \eta^{\nu\rho} G^{-1 \lambda}_{\beta}
G^{-1 \sigma}_{\gamma} F^{\beta}_{\rho\sigma}
(\nabla_{\alpha} C_{\lambda}^{\gamma}) \\
&& - \frac{1}{2} \eta^{\nu\rho} G^{-1 \lambda}_{\gamma}
G^{-1 \sigma}_{\beta} F^{\beta}_{\rho\sigma}
(\nabla_{\alpha} C_{\lambda}^{\gamma}) \\
&& - \frac{1}{4} \eta^{\lambda\rho} G^{-1 \nu}_{\beta}
G^{-1 \sigma}_{\gamma} F^{\beta}_{\rho\sigma}
(\nabla_{\alpha} C_{\lambda}^{\gamma}) \\
&& + \frac{1}{2} \eta^{\lambda\rho} G^{-1 \nu}_{\gamma}
G^{-1 \sigma}_{\beta} F^{\beta}_{\rho\sigma}
(\nabla_{\alpha} C_{\lambda}^{\gamma}) \\
&& + \frac{1}{4} \eta^{\lambda\rho} \eta^{\nu\sigma}
g_{\alpha\beta}
G^{-1 \kappa}_{\gamma} F^{\beta}_{\rho\sigma}
(G_{\lambda}^{\delta} \nabla_{\delta} C_{\kappa}^{\gamma}) \\
&& - \frac{1}{4} \eta^{\nu\rho} G^{-1 \sigma}_{\alpha}
G^{-1 \kappa}_{\gamma} F^{\mu}_{\rho\sigma}
(\nabla_{\mu} C_{\kappa}^{\gamma}) \\
&& + \frac{1}{2} \eta^{\nu\rho} G^{-1 \sigma}_{\beta}
G^{-1 \kappa}_{\gamma} F^{\beta}_{\rho\sigma}
(\nabla_{\alpha} C_{\kappa}^{\gamma}) \\
&& + \frac{1}{4} \eta^{\lambda\rho} G^{-1 \nu}_{\beta}
G^{-1 \sigma}_{\alpha}
G^{-1 \kappa}_{\gamma} F^{\beta}_{\rho\sigma}
(G_{\lambda}^{\delta} \nabla_{\delta} C_{\kappa}^{\gamma}) \\
&& - \frac{1}{2} \eta^{\lambda\rho} G^{-1 \nu}_{\alpha}
G^{-1 \sigma}_{\beta}
G^{-1 \kappa}_{\gamma} F^{\beta}_{\rho\sigma}
(G_{\lambda}^{\delta} \nabla_{\delta} C_{\kappa}^{\gamma}) \\
&& - \frac{1}{4} \eta^{\lambda\rho} \eta^{\nu\sigma}
\nabla_{\mu} (g_{\alpha \beta} C^{\mu}_{\lambda}
 F^{\beta}_{\rho\sigma}) \\
&& - \frac{1}{4} \eta^{\nu\rho}
\nabla_{\beta} (C^{\sigma}_{\lambda} G^{-1 \lambda}_{\alpha}
 F^{\beta}_{\rho\sigma}) \\
&&  + \frac{1}{2} \eta^{\nu\rho}
\nabla_{\alpha} (C^{\sigma}_{\lambda} G^{-1 \lambda}_{\beta}
 F^{\beta}_{\rho\sigma}) \\
&& + \frac{1}{4g} \eta^{\lambda\rho}
\nabla_{\mu} \lbrack ( G^{-1 \nu}_{\beta} G^{-1 \sigma}_{\alpha}
G^{\mu}_{\lambda} - \delta^{\nu}_{\beta} \delta^{\sigma}_{\alpha}
\delta^{\mu}_{\lambda}) F^{\beta}_{\rho\sigma}
\rbrack  \\
&& - \frac{1}{2 g} \eta^{\lambda\rho}
\nabla_{\mu} \lbrack ( G^{-1 \nu}_{\alpha} G^{-1 \sigma}_{\beta}
G^{\mu}_{\lambda} - \delta^{\nu}_{\alpha} \delta^{\sigma}_{\beta}
\delta^{\mu}_{\lambda}) F^{\beta}_{\rho\sigma}
\rbrack  \\
&& - \frac{1}{4} \eta^{\kappa \rho} G^{-1 \nu}_{\beta}
G^{-1 \lambda}_{\alpha} G^{-1 \sigma}_{\gamma}
F^{\beta}_{\rho\sigma} F^{\gamma}_{\kappa\lambda}  \\
&& + \frac{1}{2} \eta^{\kappa \rho} G^{-1 \nu}_{\gamma}
G^{-1 \lambda}_{\alpha} G^{-1 \sigma}_{\beta}
F^{\beta}_{\rho\sigma} F^{\gamma}_{\kappa\lambda}  \\
&& - \frac{1}{8} \eta^{\mu \rho} \eta^{\lambda \sigma}
g_{\alpha \gamma} G^{-1 \nu}_{\beta}
F^{\beta}_{\rho\sigma} F^{\gamma}_{\mu\lambda}  \\
&& - \frac{1}{16} \eta^{\mu \rho} \eta^{\lambda \sigma}
g_{\beta \gamma} G^{-1 \nu}_{\alpha}
F^{\beta}_{\rho\sigma} F^{\gamma}_{ \mu\lambda}  \\
&& - \frac{1}{8} \eta^{\mu \rho} G^{-1 \nu}_{\alpha}
G^{-1 \lambda}_{\beta} G^{-1 \sigma}_{\gamma}
F^{\beta}_{\rho\sigma} F^{\gamma}_{\mu\lambda}  \\
&& + \frac{1}{4} \eta^{\mu \rho} G^{-1 \nu}_{\alpha}
G^{-1 \lambda}_{\gamma} G^{-1 \sigma}_{\beta}
F^{\beta}_{\rho\sigma} F^{\gamma}_{\mu\lambda}.
\ea
\ee
where $\eta^{\lambda\rho}$ is the space-time metric in
curvilinear coordinate system, $\nabla_{\alpha}$ is the
covariant derivative, $g_{\alpha\beta}$ is a composite
operator defined by equation (\ref{2.7}), and
$F_{\mu\nu}^{\alpha}$ is the field strength of gravitational gauge field
which is defined by
\be \label{4.3}
F_{\mu \nu}^{ \alpha} =
G_{\mu}^{\beta} \nabla_{\beta} C_{\nu}^{ \alpha}
-G_{\nu}^{ \beta} \nabla_{\beta} C_{\mu}^{ \alpha} .
\ee
The affine connection in curvilinear coordinate system
is defined by
\be \label{4.4}
\Gamma^{\lambda}_{\mu \nu}
= \frac{1}{2} \eta^{\lambda \sigma}
\left( \partial_{\nu} \eta_{\mu \sigma}
+ \partial_{\mu} \eta_{\nu \sigma}
-\partial_{\sigma} \eta_{\mu \nu} \right).
\ee
\\

The above field equation is a little complicated. Based on this
equation, we can solve it and obtain its classical solution. We
can use Mathematica to perform these calculations. The equation
(\ref{4.2}) can be equivalently expressed in another form.
Denote
\be \label{4.5}
W_{1 \alpha}^{\nu} =
\frac{\partial {\cal L}_0}{\partial C_{\nu}^{\alpha}},
\ee
\be \label{4.6}
W_{2 \alpha}^{\mu \nu} =
\frac{\partial {\cal L}_0}{\partial \partial_{\mu} C_{\nu}^{\alpha}}.
\ee
In curvilinear coordinate system, they are given by
\be
\ba{rcl}
W_{1 \alpha}^{\nu}
&=& \frac{g}{4} \eta^{\nu\rho} \eta^{\mu\sigma}
g_{\beta\gamma} F_{\rho \sigma}^{\beta}
(\nabla_{\alpha} C_{\mu}^{\gamma}) \\
&&\\
&& + \frac{g}{4} \eta^{\nu\rho}
G^{-1 \mu}_{\beta} G^{-1 \sigma}_{\gamma}
F_{\rho \sigma}^{\beta}
(\nabla_{\alpha} C_{\mu}^{\gamma}) \\
&&\\
&& - \frac{g}{2} \eta^{\nu\rho}
G^{-1 \mu}_{\gamma} G^{-1 \sigma}_{\beta}
F_{\rho \sigma}^{\beta}
(\nabla_{\alpha} C_{\mu}^{\gamma}) \\
&&\\
&& - \frac{g}{4} \eta^{\mu\rho}
G^{-1 \nu}_{\beta} G^{-1 \sigma}_{\gamma}
F_{\rho \sigma}^{\beta}
(\nabla_{\alpha} C_{\mu}^{\gamma}) \\
&&\\
&& + \frac{g}{2} \eta^{\mu\rho}
G^{-1 \nu}_{\gamma} G^{-1 \sigma}_{\beta}
F_{\rho \sigma}^{\beta}
(\nabla_{\alpha} C_{\mu}^{\gamma}) \\
&&\\
&& + \frac{g}{8} \eta^{\lambda\rho}
\eta^{\mu\sigma} g_{\alpha\gamma}
G^{-1 \nu}_{\beta} F_{\rho \sigma}^{\beta}
F^{\gamma}_{\mu\lambda} \\
&&\\
&&
- \frac{g}{4} \eta^{\mu\rho}
G^{-1 \nu}_{\beta} G^{-1 \lambda}_{\alpha}
G^{-1 \sigma}_{\gamma}
F_{\rho \sigma}^{\beta}
F^{\gamma}_{\mu\lambda} \\
&&\\
&&
+ \frac{g}{2} \eta^{\mu\rho}
G^{-1 \nu}_{\gamma} G^{-1 \lambda}_{\alpha}
G^{-1 \sigma}_{\beta}
F_{\rho \sigma}^{\beta}
F^{\gamma}_{\mu\lambda},
\ea
\label{4.7}
\ee
 and
\be
\ba{rcl}
W_{2 \alpha}^{\mu \nu}
&=&- \frac{1}{4} \eta^{\lambda \rho}
\eta^{\nu \sigma} g_{\alpha \beta }
G^{\mu}_{\lambda} F_{\rho \sigma}^{\beta} \\
&&\\
&& + \frac{1}{4} \eta^{\nu\rho} G^{-1 \sigma}_{\alpha}
F_{\rho\sigma}^{\mu} \\
&&\\
&& - \frac{1}{4} \eta^{\lambda \rho} G^{-1 \nu}_{\beta}
G^{-1 \sigma}_{\alpha} G^{\mu}_{\lambda}
F_{\rho \sigma}^{\beta} \\
&&\\
&& + \frac{1}{2} \eta^{\lambda \rho} G^{-1 \nu}_{\alpha}
G^{-1 \sigma}_{\beta} G^{\mu}_{\lambda}
F_{\rho \sigma}^{\beta} \\
&&\\
&& - \frac{1}{2} \eta^{\nu\rho} \delta^{\mu}_{\alpha}
G^{-1 \sigma}_{\beta} F_{\rho \sigma}^{\beta}.
\ea
\label{4.8}
\ee
Define
\be \label{4.9}
W^{\nu}_{\alpha} =
-\nabla_{\mu} W_{2 \alpha}^{\mu\nu}
+ W_{1 \alpha}^{\nu}
+ g G^{-1 \nu}_{\alpha} {\cal L}_0
- g G^{-1 \lambda}_{\beta} (\nabla_{\mu} C_{\lambda}^{\beta})
  W^{\mu\nu}_{2 \alpha},
\ee
then field equation (\ref{4.1}) can be expressed by
\be \label{4.10}
W^{\nu}_{\alpha} = 0.
\ee
\\

Now, we begin to solve field equation (\ref{4.10}). In spherical
symmetric case, gravitational gauge field $C_{\mu}^{\alpha}$
is selected to be
\be \label{4.11}
g C_{\mu}^{\alpha} = \left (
\ba{cccc}
1 - \frac{1}{B(r)} &0&0& 0\\
0& 1 - \frac{1}{A(r)} &0&0  \\
0&0&0&0  \\
0&0&0&0
\ea
\right ).
\ee
In this case, matrix $G$ and $G^{-1}$ are respectively given by
\be \label{4.12}
G= (G_{\mu}^{\alpha}) = \left (
\ba{cccc}
 \frac{1}{B(r)} &0&0& 0\\
0& \frac{1}{A(r)} &0&0  \\
0&0&1&0  \\
0&0&0&1
\ea
\right ),
\ee
and
\be \label{4.13}
G^{-1}= (G^{-1\mu}_{\alpha}) = \left (
\ba{cccc}
B(r) &0&0& 0\\
0& A(r) &0&0  \\
0&0&1&0  \\
0&0&0&1
\ea
\right ).
\ee
\\

In spherical coordinate system, space-time metric is
\be \label{4.14}
\eta_{\mu\nu} = \left (
\ba{cccc}
-1 &0&0& 0\\
0& 1 &0&0  \\
0&0&r^2&0  \\
0&0&0&r^2 \sin^2 \theta
\ea
\right ).
\ee
Its inverse is
\be \label{4.15}
\eta^{\mu\nu} = \left (
\ba{cccc}
-1 &0&0& 0\\
0& 1 &0&0  \\
0&0&\frac{1}{r^2}&0  \\
0&0&0&\frac{1}{r^2 \sin^2 \theta}
\ea
\right ).
\ee
Equation (\ref{2.7}) gives the definition of $g_{\alpha\beta}$
\be \label{4.16}
g_{\alpha\beta} = \left (
\ba{cccc}
-B^2(r) &0&0& 0\\
0& A^2(r) &0&0  \\
0&0&r^2&0  \\
0&0&0&r^2 \sin^2 \theta
\ea
\right ).
\ee
Affine connection is defined by equation (\ref{4.4}). In spherical
coordinate system, its only nonvanishing components are
\be \label{4.17}
 \left \lbrace
\ba{l}
\Gamma^{r}_{\theta\theta} = -r \\
\Gamma^{r}_{\varphi\varphi} = -r \sin^2 \theta \\
\Gamma^{\theta}_{r \theta} = \Gamma^{\theta}_{\theta r} = \frac{1}{r} \\
\Gamma^{\theta}_{\varphi\varphi} = - \sin \theta \cos \theta \\
\Gamma^{\varphi}_{r \varphi} = \Gamma^{\varphi}_{\varphi r} = \frac{1}{r} \\
\Gamma^{\varphi}_{\theta \varphi} = \Gamma^{\varphi}_{\varphi \theta} = { \rm ctg} \theta
\ea
\right. .
\ee
\\

Field strength of gravitational gauge field $F_{\mu\nu}^{\alpha}$
is defined by equation (\ref{4.3}). In the present case, its
only nonvanishing components are
\be \label{4.18}
 \left \lbrace
\ba{l}
-g F^{t}_{t r } = g F^{t}_{r t} = \frac{B'(r)}{A(r) B^2(r)} \\
-g F^{\theta}_{r \theta} =  g F^{\theta}_{ \theta r}
= \frac{A(r) -1}{r A(r)} \\
-g F^{\varphi}_{r \varphi} =  g F^{\varphi}_{ \varphi r}
= \frac{A(r) -1}{r A(r)}
\ea
\right. .
\ee
In above equations, a prime means differentiation with respect to $r$.
\\

Using all above results, we can calculate $W^{\nu}_{\alpha}$. It is found
that, its only nonvanishing components are
\be \label{4.19}
W^t_t =  \frac{B(r) [-A(r) + A^3(r) + 2 r A'(r) ]}
{2 g r^2 A^3(r)},
\ee
\be \label{4.20}
W^r_r =  \frac{ B(r) [-1 + A^2(r)] - 2 r B'(r) }
{2 g r^2 A(r) B(r)},
\ee
and
\be \label{4.21}
W^{\theta}_{\theta} = W_{\varphi}^{\varphi} =
\frac{ A'(r) [B(r) + r B'(r)]  - A(r)[ B'(r) + r B''(r) ]}
{2 g  A^3(r) B(r)}.
\ee
Then field equation (\ref{4.10}) gives the following three
equations
\be \label{4.22}
-A(r) + A^3(r) + 2 r A'(r) =0,
\ee
\be \label{4.23}
 B(r)  - A^2(r) B(r) + 2 r B'(r) =0,
\ee
and
\be \label{4.24}
 A'(r) [B(r) + r B'(r)]  - A(r)[ B'(R) + r B''(r) ]=0.
\ee
From equations (\ref{4.22}) and (\ref{4.23}) we have
\be \label{4.25}
\frac{\rm d}{{\rm d} r}( A(r) B(r)) =0.
\ee
It gives
\be \label{4.26}
 A(r) B(r)= c_0,
\ee
where $c_0$ is a constant. When $r$ approaches infinity,
this is no gravity, and gravitational
gauge field $C_{\mu}^{\alpha}$ vanish. In this case, $A(r)=B(r)=1$.
So, $c_0=1$, and equation (\ref{4.26}) becomes
\be \label{4.27}
 A(r) B(r)= 1,
\ee
Equation (\ref{4.22}) gives
\be \label{4.28}
\frac{\rm d}{{\rm d} r}\frac{ r }{A^2(r) } =1.
\ee
From equations (\ref{4.27}) and (\ref{4.28}), we obtain final
solution
\be \label{4.29}
 A(r) = \frac{1}{\sqrt{1-\frac{2 G M}{r}}},
\ee
\be \label{4.30}
 B(r) = \sqrt{1-\frac{2 G M}{r}}.
\ee
Then  (\ref{4.16}) gives the following results
\be \label{4.31}
g_{\alpha\beta} = \left (
\ba{cccc}
-(1-\frac{2 G M}{r}) &0&0& 0\\
0& \frac{1}{1-\frac{2 G M}{r}} &0&0  \\
0&0&r^2&0  \\
0&0&0&r^2 \sin^2 \theta
\ea
\right ),
\ee
which is just the Schwarzschild solution in general relativity.
Therefor, solutions (\ref{4.29}) and (\ref{4.30}) correspond
to the Schwarzschild solution. After we know $A(r)$ and $B(r)$,
we can calculate magnitude of gravitational gauge field from
equation (\ref{4.11}). The only nonvanishing components of gravitational
gauge field are
\be \label{4.32}
 g C^t_t  = 1- \frac{1}{\sqrt{1-\frac{2 G M}{r}}},
\ee
\be \label{4.23}
 g C^r_r  = 1- \sqrt{1-\frac{2 G M}{r}}.
\ee
\\

\section{Equation of Motion of a Mass Point in Gravitational
            Gauge Field}
\setcounter{equation}{0}

General relativity is a theory of space-time geometry. In general
relativity, the equation of motion of a mass point is the
geodesic equation, for geodesic curve is the path which has
shortest length between two points.
So, in general relativity,  selecting the geodesic equation
as the equation of motion of a mass point is consistent with basic
spirit of geometry. Gauge theory of gravity is not a theory of
space-time geometry. In gauge theory of gravity, space-time
is always flat, and geodesic line is always a straight line
in Minkowski space-time, which is the orbit of an inertial motion.
When there is gravitational force on the mass point, its orbit
deviates from the geodesic line.
So, we can not select the geodesic
equation as the equation of motion of a mass point in
gravitational field. How to set up the equation of motion
of a mass point in a logically self-consistent way? \\

As we have stated before, the basic viewpoint in gauge theory
of gravity is that gravity is a kind of fundamental interactions
in flat Minkowski space-time. For a classical mass point in
gravitational field, it will feel a gravitational force acting
on it. Denote the  force as $f^{\mu}$, then according to
Newton's second law of motion, the equation
of motion of a mass point should be
\be \label{5.1}
 \frac{{\rm d} P^{\mu}}{{\rm d} \tau} = f^{\mu},
\ee
where $P^{\mu}$ is the canonical momentum of the mass point.
Now, our central task is to determine the form of the
gravitational force $f^{\mu}$ on the mass point.\\

In Newton's theory of gravity, the gravitational force on
a mass point is
\be \label{5.2}
 \svec{f}= - \frac{G M m}{r^2} \hat{r},
\ee
where $G$ is the Newtonian gravitational constant,
$M$ is the mass of gravitational source, $m$ is the mass
of the mass point, $\hat r=\frac{\svec{r}}{r}$ is a unit vector,
and $r$ is the distance between the mass point and the
center of mass of the source. But this
form is not relativistic, we can not directly use it in the
equation (\ref{5.1}). \\

Ir order to determine the form of gravitational force $f^{\mu}$,
we need first know its basic properties.
The basic requirements of the gravitational force $f^{\mu}$ are
the following three:

\begin{enumerate}
\item under Lorentz transformations, the gravitational force $f^{\mu}$
is a Lorentz 4-vector;
\item under gravitational gauge transformations, the gravitational
force $f^{\mu}$   transforms covariantly
\be \label{5.3}
 f^{\mu} \to (\ehat f^{\mu} );
\ee
\item in non-relativistic limit, the gravitational force $f^{\mu}$
returns to Newton's formula (\ref{5.2}).
\end{enumerate}

In non-relativistic case, the gravitational field generated by
a mass $M$ is \cite{05,06,07,08}
\be \label{5.4}
g C^{0}_{0} = - \frac{G M}{r}.
\ee
Therefor, in leading order approximation, the corresponding
field strength is
\be \label{5.5}
g F^{0}_{i0} =  \frac{G M}{r^2} \hat r_i.
\ee
Then equation (\ref{5.2}) can be changed into
\be \label{5.6}
 f_i = - g F^{0}_{i0} m.
\ee
For non-relativistic case, we have
\be \label{5.7}
 P^0 = m, ~~~ \frac{{\rm d} x_0}{{\rm d} \tau}= - \frac{{\rm d} x^0}{{\rm d} \tau} =-1.
\ee
So, equation (\ref{5.6}) becomes
\be \label{5.8}
 f_i =  g F^{0}_{i0} \frac{{\rm d} x_0}{{\rm d} \tau} P^0.
\ee
The simplest way to generate the above formula to relativistic
form is to replace index 0 to index $\lambda$ or $\alpha$ and
sum over repeated indexes. So, we have
\be \label{5.9}
 f_i =  g F^{\alpha}_{i\lambda}
 \frac{{\rm d} x_{\alpha}}{{\rm d} \tau} P^{\lambda}.
\ee
Therefore, the gravitational force $f^{\mu}$ should be
\be \label{5.10}
 f^{\mu} =  g \eta^{\mu\nu} F^{\alpha}_{\nu \lambda}
 \frac{{\rm d} x_{\alpha}}{{\rm d} \tau} P^{\lambda}.
\ee
In literature \cite{23}, we have obtained similar result by
using a quite different method.
It is found that this formula indeed satisfies the three
requirements above. So, the $f^{\mu}$ given by equation
(\ref{5.10}) is the gravitational force on the mass point.
Then equation of motion of the mass point in gravitational
field is
\be \label{5.11}
 \frac{{\rm d} P^{\mu}}{{\rm d} \tau} =
 g \eta^{\mu\nu} F^{\alpha}_{\nu \lambda}
 \frac{{\rm d} x_{\alpha}}{{\rm d} \tau} P^{\lambda}.
\ee
In gauge theory of gravity, the definition of
$\frac{{\rm d} x_{\alpha}}{{\rm d} \tau}$ is given by
\be \label{5.12}
 \frac{{\rm d} x_{\alpha}}{{\rm d} \tau}
 = g_{\alpha \beta} \frac{{\rm d} x^{\beta}}{{\rm d} \tau},
\ee
so, equation (\ref{5.11}) becomes
\be \label{5.13}
 \frac{{\rm d} P^{\mu}}{{\rm d} \tau} =
 g \eta^{\mu\nu} g_{\alpha \beta} F^{\alpha}_{\nu \lambda}
 \frac{{\rm d} x^{\beta}}{{\rm d} \tau} P^{\lambda}.
\ee
\\

In quantum gauge theory of gravity, energy-momentum operator
$p_{\alpha}$ is given by
\be \label{5.14}
p_{\alpha} = - i \partial_{\alpha}.
\ee
According to gauge principle, gauge canonical momentum
$P_{\mu}$ is
\be \label{5.15}
P_{\mu} = - i D_{\mu}.
\ee
Equation (\ref{2.3}) gives  their relation
\be \label{5.16}
P_{\mu} = G_{\mu}^{\alpha} p_{\alpha}.
\ee
Using equation (\ref{2.7}) and
\be \label{5.17}
P_{\mu} = \eta_{\mu\nu} P^{\nu}, ~~~
p_{\alpha} = g_{\alpha\beta} p^{\beta},
\ee
we can change equation (\ref{5.16}) into
\be \label{5.18}
P^{\mu} = G^{-1 \mu}_{\gamma} p^{\gamma}.
\ee
Multiply both side of equation (\ref{5.13}) with
$G_{\mu}^{\gamma}$ and sum over index $\mu$, we get
\be \label{5.19}
G_{\mu}^{\gamma} \frac{{\rm d} (G^{-1 \mu}_{\gamma_1} p^{\gamma_1})}
{{\rm d} \tau} =
 g \eta^{\mu\nu} G_{\mu}^{\gamma}
 g_{\alpha_1 \beta} F^{\alpha_1}_{\nu \lambda}
 \frac{{\rm d} x^{\beta}}{{\rm d} \tau} ( G^{-1 \lambda}_{\alpha} p^{\alpha}).
\ee
Left hand side of equation (\ref{5.19}) gives
\be \label{5.20}
\ba{rl}
& \frac{{\rm d}  p^{\gamma}} {{\rm d} \tau}
+ G_{\mu}^{\gamma}
(\frac{\rm d}{{\rm d} \tau} G^{-1 \mu}_{\beta} ) p^{\beta} \\
=& \frac{{\rm d}  p^{\gamma}} {{\rm d} \tau}
+ G_{\mu}^{\gamma} (\partial_{\alpha} G^{-1 \mu}_{\beta} )
\frac{{\rm d}x^{\alpha}}{{\rm d} \tau}   p^{\beta}.
\ea
\ee
Because $\frac{{\rm d}x^{\alpha}}{{\rm d} \tau}   p^{\beta}$
is symmetric under exchange indexes $\alpha$ and $\beta$,
it can be further changed into
\be \label{5.21}
\frac{{\rm d}  p^{\gamma}} {{\rm d} \tau}
+ \bar\Gamma^{\gamma}_{\alpha\beta}
\frac{{\rm d}x^{\alpha}}{{\rm d} \tau}   p^{\beta},
\ee
where
\be \label{5.22}
\ba{rcl}
\bar\Gamma^{\gamma}_{\alpha\beta} &=&
 \frac{1} {2}  G_{\mu}^{\gamma}
(\partial_{\alpha} G^{-1 \mu}_{\beta} +
\partial_{\beta} G^{-1 \mu}_{\alpha}) \\
&=& - \frac{1}{2}
( G^{-1 \mu}_{\beta} \partial_{\alpha} G_{\mu}^{\gamma} +
G^{-1 \mu}_{\alpha} \partial_{\beta} G_{\mu}^{\gamma}).
\ea
\ee
So, equation (\ref{5.19}) was changed into the following form
\be \label{5.23}
\frac{{\rm d}  p^{\gamma}} {{\rm d} \tau}
+ \bar\Gamma^{\gamma}_{\alpha\beta}
\frac{{\rm d}x^{\alpha}}{{\rm d} \tau}   p^{\beta}=
\frac{g}{2} \eta^{\mu\nu} G_{\mu}^{\gamma}
\left ( g_{\alpha_1 \beta} G^{-1 \lambda}_{\alpha} +
g_{\alpha_1 \alpha} G^{-1 \lambda}_{\beta} \right)
 F^{\alpha_1}_{\nu \lambda}
 \frac{{\rm d} x^{\alpha}}{{\rm d} \tau}  p^{\beta}.
\ee
This is the equation of motion of a mass point in gravitational field.
\\

\section{Equation of Motion in Spherical Coordinate System}
\setcounter{equation}{0}

In above chapter, we obtain a equation of motion of a classical mass
point which is moving in gravitational field. Equation (\ref{5.23})
is only valid in Descartes coordinate system. According to
the discussions in chapter 3, in arbitrary curvilinear
coordinate system, equation (\ref{5.23}) should be changed
into
\be \label{6.1}
\frac{D  p^{\gamma}} {D \tau}
+ \bar\Gamma^{\gamma}_{\alpha\beta}
\frac{{\rm d}x^{\alpha}}{{\rm d} \tau}   p^{\beta}=
\frac{g}{2} \eta^{\mu\nu} G_{\mu}^{\gamma}
\left ( g_{\alpha_1 \beta} G^{-1 \lambda}_{\alpha} +
g_{\alpha_1 \alpha} G^{-1 \lambda}_{\beta} \right)
 F^{\alpha_1}_{\nu \lambda}
 \frac{{\rm d} x^{\alpha}}{{\rm d} \tau}  p^{\beta},
\ee
where
\be \label{6.2}
\frac{D  p^{\gamma}} {D \tau} =
\frac{{\rm d} p^{\gamma}}{{\rm d} \tau}
+ \Gamma^{\gamma}_{\alpha\beta}
\frac{{\rm d} x^{\alpha}}{{\rm d} \tau}  p^{\beta},
 \ee
with $\Gamma^{\gamma}_{\alpha\beta}$ is the affine connection
in curvilinear coordinate system defined by (\ref{4.4}), and
\be \label{6.3}
\ba{rcl}
\bar\Gamma^{\gamma}_{\alpha\beta} &= &
\frac{1}{2}   G_{\mu}^{\gamma}
(\nabla_{\alpha} G^{-1 \mu}_{\beta} +
\nabla_{\beta} G^{-1 \mu}_{\alpha}) \\
&=&  \frac{1}{2}  G_{\mu}^{\gamma} \lbrack
\partial_{\alpha} G^{-1 \mu}_{\beta}
+ \Gamma^{\mu}_{\alpha\nu} G^{-1 \nu}_{\beta}
- \Gamma^{\gamma}_{\alpha\beta} G^{-1 \mu}_{\gamma}  \\
&&~~~+ \partial_{\beta} G^{-1 \mu}_{\alpha}
+ \Gamma^{\mu}_{\beta\nu} G^{-1 \nu}_{\alpha}
- \Gamma^{\gamma}_{\beta\alpha} G^{-1 \mu}_{\gamma} \rbrack .
\ea
\ee
\\

In spherical coordinate system, gravitational gauge field
$C_{\mu}^{\alpha}$ is given by (\ref{4.11}), matrixes $G$
and $G^{-1}$ are given by (\ref{4.12}) and (\ref{4.13})
respectively, metrics $\eta_{\mu\nu}$ and $\eta^{\mu\nu}$
are given by (\ref{4.14}) and (\ref{4.15}), $g_{\alpha\beta}$
is given by (\ref{4.16}), affine connection
$\Gamma^{\gamma}_{\alpha\beta}$ is given by (\ref{4.17}),
and field strength $F^{\alpha}_{\mu\nu}$ is given by
(\ref{4.18}). $\bar\Gamma^{\gamma}_{\alpha\beta}$ is calculate
from equation (\ref{6.3}). Its only nonvanishing components are
\be \label{6.4}
 \left \lbrace
\ba{l}
\bar\Gamma^{t}_{tr} = \bar\Gamma^{t}_{rt} =
\frac{B'(r)}{ 2 B(r)} \\
\bar\Gamma^{r}_{r r} = \frac{A'(r)}{A(r)} \\
\bar\Gamma^{r}_{\theta \theta}  = r( 1- \frac{1}{A(r)} ) \\
\bar\Gamma^{r}_{\varphi\varphi}
=  r \sin^2 \theta ( 1- \frac{1}{A(r)} ) \\
\bar\Gamma^{\theta}_{r \theta} = \bar\Gamma^{\theta}_{\theta r}
= \frac{A(r) -1}{2 r} \\
\bar\Gamma^{\varphi}_{r  \varphi} = \Gamma^{\varphi}_{\varphi r}
= \frac{A(r) - 1 }{2 r}
\ea
\right. .
\ee
Denote
\be \label{6.5}
f^{\gamma}_{\alpha\beta} =
 - \frac{g}{2} \eta^{\mu\nu} G_{\mu}^{\gamma}
\left ( g_{\alpha_1 \beta} G^{-1 \lambda}_{\alpha} +
g_{\alpha_1 \alpha} G^{-1 \lambda}_{\beta} \right)
 F^{\alpha_1}_{\nu \lambda},
\ee
then equation (\ref{6.1}) is changed into a much simpler form
\be \label{6.6}
\frac{D  p^{\gamma}} {D \tau}
+ ( \bar\Gamma^{\gamma}_{\alpha\beta} + f^{\gamma}_{\alpha\beta} )
\frac{{\rm d}x^{\alpha}}{{\rm d} \tau}   p^{\beta}= 0.
\ee
The only nonvanishing components of $f^{\gamma}_{\alpha\beta}$
are
\be \label{6.7}
 \left \lbrace
\ba{l}
f^{t}_{tr} = f^{t}_{rt} =
\frac{B'(r)}{ 2 B(r)} \\
f^{r}_{t t} = \frac{B(r) B'(r)}{A^2(r)} \\
f^{r}_{\theta \theta}  =  \frac{A(r) - 1}{A^2(r)} \cdot r \\
f^{r}_{\varphi\varphi}
=  \frac{A(r) - 1}{A^2(r)} \cdot r \sin^2 \theta  \\
f^{\theta}_{r \theta} = f^{\theta}_{\theta r}
= \frac{1- A(r) }{2 r} \\
f^{\varphi}_{r  \varphi} = f^{\varphi}_{\varphi r}
= \frac{1- A(r) }{2 r}
\ea
\right. .
\ee
\\

Using nonvanishing components of affine connection
$\Gamma^{\gamma}_{\alpha\beta}$ given by (\ref{4.17}),
$\bar\Gamma^{\gamma}_{\alpha\beta}$ given by (\ref{6.4}),
and $f^{\gamma}_{\alpha\beta}$ given by (\ref{6.7}),
we find from (\ref{6.6}) that
\be \label{6.8}
\frac{{\rm d} p^t}{{\rm d} \tau}
+ 2 \frac{B'(r)}{B(r)} \frac{{\rm d} r}{{\rm d}\tau} p^t = 0,
\ee
\be \label{6.9}
\frac{{\rm d} p^r}{{\rm d} \tau}
+  \frac{B(r) B'(r)}{A^2(r)} \frac{{\rm d} t}{{\rm d}\tau} p^t
+  \frac{A'(r)}{A(r)} \frac{{\rm d} r}{{\rm d}\tau} p^r
-  \frac{r}{A^2(r)} \frac{{\rm d} \theta}{{\rm d}\tau} p^{\theta}
-  \frac{r \sin^2 \theta}{A^2(r)} \frac{{\rm d} \varphi}{{\rm d}\tau} p^{\varphi}
= 0,
\ee
\be \label{6.10}
\frac{{\rm d} p^{\theta}}{{\rm d} \tau}
+  \frac{2}{r} \frac{{\rm d} r}{{\rm d}\tau} p^{\theta}
-  \sin\theta \cos\theta \frac{{\rm d} \varphi}{{\rm d}\tau} p^{\varphi}
= 0,
\ee
\be \label{6.11}
\frac{{\rm d} p^{\varphi}}{{\rm d} \tau}
+  \frac{2}{r} \frac{{\rm d} r}{{\rm d}\tau} p^{\varphi}
+ 2 {\rm ctg}\theta \frac{{\rm d} \theta}{{\rm d}\tau} p^{\varphi}
= 0.
\ee
Since the field is isotropic, we may consider the orbit of the
mass point to be confined to the equatorial plane, that is, in
the $\theta = \frac{\pi}{2}$ plane. In this case, equation
(\ref{6.10}) vanishes, other three equations are changed into
\be \label{6.12}
\frac{{\rm d} p^t}{{\rm d} \tau}
+ 2 \frac{B'(r)}{B(r)} \frac{{\rm d} r}{{\rm d}\tau} p^t = 0,
\ee
\be \label{6.13}
\frac{{\rm d} p^r}{{\rm d} \tau}
+  \frac{B(r) B'(r)}{A^2(r)} \frac{{\rm d} t}{{\rm d}\tau} p^t
+  \frac{A'(r)}{A(r)} \frac{{\rm d} r}{{\rm d}\tau} p^r
-  \frac{r }{A^2(r)} \frac{{\rm d} \varphi}{{\rm d}\tau} p^{\varphi}
= 0,
\ee
\be \label{6.14}
\frac{{\rm d} p^{\varphi}}{{\rm d} \tau}
+  \frac{2}{r} \frac{{\rm d} r}{{\rm d}\tau} p^{\varphi}
= 0.
\ee
When we discuss equation of motion of photon, we should
use trajectory parameter $s$ instead of proper time $\tau$
in above equations. We can also consider photon as limit that
the rest mass $m$ approaches zero, and use all above equations
to calculate its motion in gravitational field.  \\

In order to solve these equations, we first look for constants
of the motion. Equation (\ref{6.12}) gives
\be \label{6.15}
\frac{{\rm d} }{{\rm d} \tau} \left (
{\rm ln} p^t + {\rm ln} B^2(r) \right )
 = 0.
\ee
So,
\be \label{6.16}
 p^t  B^2(r) = E_0 ,
\ee
where $E_0$ is a constant. $p^t$ can be expressed in terms of $B(r)$
\be \label{6.17}
 p^t   = \frac{E_0}{B^2(r)}.
\ee
Equation (\ref{6.14}) can be changed into
\be \label{6.18}
\frac{{\rm d} }{{\rm d} \tau} \left (
{\rm ln} p^{\varphi} + {\rm ln} r^2 \right )
 = 0,
\ee
which gives
\be \label{6.19}
 r^2 p^{\varphi} = J ,
\ee
where $J$ is an another constant. $J$ plays the role of angular
momentum. Momentum $p^{\varphi}$ is
\be \label{6.20}
 p^{\varphi} = \frac{J}{r^2} .
\ee
\\

Inserting equations (\ref{6.17}) and (\ref{6.20}) into (\ref{6.13})
gives the following equation of motion
\be \label{6.21}
\frac{{\rm d} p^r}{{\rm d} \tau}
+  \frac{E_0 B'(r)}{A^2(r) B(r)} \frac{{\rm d} t}{{\rm d}\tau}
+  \frac{A'(r)}{A(r)} \frac{{\rm d} r}{{\rm d}\tau} p^r
-  \frac{J }{r A^2(r)} \frac{{\rm d} \varphi}{{\rm d}\tau}
= 0.
\ee
Multiply both side of the above equation with $2 A^2(r) p^r$, we get
\be \label{6.22}
2 A^2(r) p^r \frac{{\rm d} p^r}{{\rm d} \tau}
+  \frac{2 E_0 B'(r)}{ B(r)}  p^r \frac{{\rm d} t}{{\rm d}\tau}
+  2 A(r) A'(r) (p^r)^2 \frac{{\rm d} r}{{\rm d}\tau}
-  \frac{2 J }{r }  p^r \frac{{\rm d} \varphi}{{\rm d}\tau}
= 0,
\ee
which can be written into another form
\be \label{6.23}
\frac{{\rm d} }{{\rm d} \tau}\left (  A^2(r) (p^r)^2 \right )
+  \frac{2 E_0 B'(r)}{ B(r)}  p^r \frac{{\rm d} t}{{\rm d}\tau}
-  \frac{2 J }{r }  p^r \frac{{\rm d} \varphi}{{\rm d}\tau}
= 0.
\ee
Using the following two relations
\be \label{6.24}
p^r \frac{{\rm d}t}{{\rm d}\tau}=
p^t \frac{{\rm d}r}{{\rm d}\tau},
\ee
and
\be \label{6.25}
p^r \frac{{\rm d}\varphi}{{\rm d}\tau}=
p^{\varphi} \frac{{\rm d}r}{{\rm d}\tau},
\ee
we can change equation (\ref{6.23}) into
\be \label{6.26}
\frac{{\rm d} }{{\rm d} \tau}\left (  A^2(r) (p^r)^2 \right )
+  \frac{2 E_0 B'(r)}{ B(r)}  p^t \frac{{\rm d} r}{{\rm d}\tau}
-  \frac{2 J }{r }  p^{\varphi} \frac{{\rm d} r}{{\rm d}\tau}
= 0.
\ee
Inserting equations (\ref{6.17}) and (\ref{6.20}) into (\ref{6.26})
gives
\be \label{6.27}
\frac{{\rm d} }{{\rm d} \tau} \left \lbrack
 A^2(r) (p^r)^2 - \frac{E_0^2 }{ B^2(r)}
 + \frac{J^2 }{r^2}  \right \rbrack = 0.
\ee
So,
\be \label{6.28}
 A^2(r) (p^r)^2 - \frac{E_0^2 }{ B^2(r)}
 + \frac{J^2 }{r^2}  = E,
\ee
where $E$ is a constant of the motion. Equations
(\ref{6.17}), (\ref{6.20}) and  (\ref{6.28}) will be
used as the basic equations of motion to calculate
classical tests of gauge theory of  gravity. This set of equations
do not contain proper time $\tau$ in their expressions,
so they are also directly applicable to photons.
\\

\section{The Deflection of Light by the Sun}
\setcounter{equation}{0}

Using the following relations
\be \label{7.1}
 p^r = \frac{p^r}{p^{\varphi}} p^{\varphi}
 = p^{\varphi} \frac{{\rm d} r}{{\rm d} \varphi}
 = \frac{J}{r^2} \frac{{\rm d} r}{{\rm d} \varphi},
\ee
equation (\ref{6.28}) can be changed into
\be \label{7.2}
 \frac{ J^2 A^2(r)}{r^4}  \left (
 \frac{{\rm d} r}{{\rm d} \varphi} \right )^2
 + \frac{J^2 }{r^2} - \frac{E_0^2 }{ B^2(r)}  = E.
\ee
At distance $r_0$ of the closest approach to the sun,
$\frac{{\rm d}r}{{\rm d}\varphi}$ vanishes, so equation
(\ref{7.2}) gives
\be \label{7.3}
 \frac{J^2 }{r_0^2} - \frac{E_0^2 }{ B^2(r_0)}  = E.
\ee
Consider photon approaching the sun from very great distance.
At infinity, the gravitational gauge field $C_{\mu}^{\alpha}$
vanishes, and
\be \label{7.4}
A(\infty) = B(\infty) = 1.
\ee
So, at infinity, equation (\ref{6.28}) gives
\be \label{7.5}
 E = (p^r)^2 - E_0^2  .
\ee
For a photon, when there is no gravity at infinity and the photon
is moving towards the sun, its radial momentum $p^r$ equals
its energy $p^t$
\be \label{7.6}
p^r = p^t =  E_0  .
\ee
Then equation (\ref{7.5}) gives
\be \label{7.7}
 E = 0  .
\ee
Using (\ref{7.7}) in (\ref{7.3}) gives
\be \label{7.8}
 J =   \frac{r_0 E_0 }{ B(r_0)} .
\ee
Inserting equations (\ref{7.7}) and (\ref{7.8}) into
(\ref{7.2}) gives
\be \label{7.9}
 \frac{  A^2(r)}{r^4}  \left (
 \frac{{\rm d} r}{{\rm d} \varphi} \right )^2
 + \frac{1 }{r^2} = \frac{B^2(r_0) }{ r_0^2 B^2(r)}  .
\ee
After some simple calculations from equation (\ref{7.9}),
we can change its form into
\be \label{7.10}
{\rm d} \varphi = \frac{A(r)}
{ \sqrt{\frac{B^2(r_0)}{B^2(r)}
       \left ( \frac{r}{r_0} \right )^2  -1}}
\frac{{\rm d}r }{r}  .
\ee
\\

Using solution (\ref{4.30}), in first order approximation, we have
\be \label{7.11}
\frac{B^2(r_0)}{B^2(r)}
       \left ( \frac{r}{r_0} \right )^2  -1
= \left \lbrack \left ( \frac{r}{r_0}
\right )^2 -1 \right \rbrack
\left \lbrack 1 -
\frac{2 G M r}{r_0 (r + r_0)} + \cdots
\right \rbrack,
\ee
where $\cdots$ represents higher order approximation. Integrate
(\ref{7.10}) from infinity to distance $r$, we get
\be \label{7.12}
\varphi(r) - \varphi(\infty) = \int_r^{\infty}
A \left \lbrack 1 -
\frac{2 G M r}{r_0 (r + r_0)} + \cdots
\right \rbrack ^{- \frac{1}{2}}
\frac{{\rm d}r /r}{ \sqrt{ \left ( \frac{r}{r_0}
\right )^2 -1}}.
\ee
Inserting (\ref{4.29}) into above equation gives
\be \label{7.13}
\varphi(r) - \varphi(\infty) = \int_r^{\infty}
 \left \lbrack 1 + \frac{G M}{r}
+ \frac{ G M r}{r_0 (r + r_0)} + \cdots
\right \rbrack
\frac{{\rm d}r }{r  \sqrt{ \left ( \frac{r}{r_0}
\right )^2 -1}}.
\ee
\\

Using the following integration formula
\be \label{7.14}
 \int \frac{{\rm d} x}{ x \sqrt{x^2-1}}
 = \cos^{-1} \frac{1}{x},
\ee
\be \label{7.15}
 \int \frac{{\rm d} x}{ x^2 \sqrt{x^2-1}}
 = \frac{\sqrt{x^2-1}}{x},
\ee
and
\be \label{7.16}
 \int \frac{{\rm d} x}{ x \sqrt{x^2-2 x}}
 = \sqrt{1 - \frac{2}{x}},
\ee
we have
\be \label{7.17}
\ba{rcl}
 \int_r^{\infty}
\frac{{\rm d}r }{r  \sqrt{ \left ( \frac{r}{r_0}
\right )^2 -1}} &=&
\int_r^{\infty}
\frac{{\rm d} \left ( \frac{r}{r_0} \right )}
{  \left ( \frac{r}{r_0} \right ) \sqrt{ \left ( \frac{r}{r_0}
\right )^2 -1}} \\
&=& \left . \cos^{-1} \frac{1}{x} \right |_{r/r_0}^{\infty} \\
&=& \frac{\pi}{2} - \cos^{-1} \left ( \frac{r_0}{r} \right ),
\ea
\ee
\be \label{7.18}
\ba{rcl}
 \int_r^{\infty} \frac{G M}{r}
\frac{{\rm d}r }{r  \sqrt{ \left ( \frac{r}{r_0}
\right )^2 -1}} &=&
\frac{G M}{r_0} \int_r^{\infty}
\frac{{\rm d} \left ( \frac{r}{r_0} \right )}
{  \left ( \frac{r}{r_0} \right )^2 \sqrt{ \left ( \frac{r}{r_0}
\right )^2 -1}} \\
&=& \frac{G M}{r_0} \left . \frac{\sqrt{x^2-1}}{x}
\right  |_{r/r_0}^{\infty} \\
&=&  \frac{G M}{r_0} \left ( 1 - \sqrt{1-
\left (\frac{r_0}{r} \right )^2 } \right ),
\ea
\ee
and
\be \label{7.19}
\ba{rcl}
 \int_r^{\infty} \frac{G M r}{r_0 (r + r_0)}
\frac{{\rm d}r }{r  \sqrt{ \left ( \frac{r}{r_0}
\right )^2 -1}} &=&
\frac{G M}{r_0} \int_r^{\infty}
\frac{{\rm d} \left ( \frac{r}{r_0}  +1 \right )}
{  \left ( \frac{r}{r_0} +1 \right )
\sqrt{ \left ( \frac{r}{r_0} + 1
\right )^2 -2  \left ( \frac{r}{r_0} + 1 \right )}} \\
&=& \frac{G M}{r_0} \left . \sqrt{1- \frac{2}{x}}
\right  |_{\frac{r}{r_0}+1}^{\infty} \\
&=&  \frac{G M}{r_0} \left ( 1 - \sqrt{\frac{r-r_0}{r+r_0}
} \right ).
\ea
\ee
Inserting (\ref{7.17}), (\ref{7.18}) and (\ref{7.19})
into (\ref{7.13}), we have
\be \label{7.20}
\varphi(r) - \varphi(\infty) =
\frac{\pi}{2} - \cos^{-1} \left ( \frac{r_0}{r} \right )
+ \frac{G M}{r_0} \left ( 2 - \sqrt{1-
\left (\frac{r_0}{r} \right )^2}
-\sqrt{\frac{r-r_0}{r+r_0}}
\right ).
\ee
The deflection of the photon orbit from a straight line is
\be \label{7.21}
\Delta\varphi = 2 | \varphi(r_0) - \varphi(\infty) | - \pi
= \frac{4 G M}{r_0}.
\ee
This result is the same as that in general relativity.
\\

\section{The Precession of the Perihelia of the Orbits of
            the Inner planets}
\setcounter{equation}{0}

Consider a inner planet bound in an elliptical orbit around
the sun. At perihelia and aphelia, $r$ reaches its minimum and
maximum values $r_-$ and $r_+$. At  $r_-$ and $r_+$,
$\frac{{\rm d}r}{{\rm d}\varphi}$ vanishes
\be \label{8.1}
\frac{{\rm d}r}{{\rm d}\varphi} =0.
\ee
So, equation (\ref{7.2}) gives
\be \label{8.2}
 \frac{J^2 }{r_+^2} - \frac{E_0^2 }{ B^2(r_+)}  = E,
\ee
and
\be \label{8.3}
 \frac{J^2 }{r_-^2} - \frac{E_0^2 }{ B^2(r_-)}  = E.
\ee
Equations (\ref{8.2}) and (\ref{8.3}) can be considered as
equations of $J$ and $E$. From these two equations we
can derive values of $J$ and $E$
\be \label{8.4}
J^2 = \frac{ E_0^2 \left ( \frac{1}{B^2(r_-)}
- \frac{1}{B^2(r_+)} \right )}
{\frac{1}{r_-^2} - \frac{1}{r_+^2}},
\ee
\be \label{8.5}
E = \frac{ E_0^2 \left ( \frac{r_-^2}{B^2(r_-)}
- \frac{r_+^2}{B^2(r_+)} \right )}
{r_+^2 - r_-^2}.
\ee
\\

From equation (\ref{7.2}), we get
\be \label{8.6}
{\rm d}\varphi = \frac{A(r)}{r^2}
\left ( \frac{E}{J^2}
+ \frac{E_0^2}{J^2 B^2(r)}
- \frac{1}{r^2}
\right )^{- \frac{1}{2}} {\rm d}r .
\ee
Using (\ref{8.4}) and (\ref{8.5}), we have
\be \label{8.7}
\frac{E}{J^2}
+ \frac{E_0^2}{J^2 B^2(r)}
= \frac{ r_-^2 \left ( \frac{1}{B^2(r_-)}
- \frac{1}{B^2(r)}
\right )
- r_+^2 \left ( \frac{1}{B^2(r_+)}
- \frac{1}{B^2(r)}
\right )}
{r_+^2 r_-^2  \left ( \frac{1}{B^2(r_-)}
- \frac{1}{B^2(r_+)}
\right )}.
\ee
Define
\be \label{8.8}
\ba{rcl}
I(r) &=&  \frac{E}{J^2}
+ \frac{E_0^2}{J^2 B^2(r)}
- \frac{1}{r^2} \\
&=&  \frac{ r_-^2 \left ( \frac{1}{B^2(r_-)}
- \frac{1}{B^2(r)}
\right )
- r_+^2 \left ( \frac{1}{B^2(r_+)}
- \frac{1}{B^2(r)}
\right )}
{r_+^2 r_-^2  \left ( \frac{1}{B^2(r_-)}
- \frac{1}{B^2(r_+)}
\right )}
- \frac{1}{r^2}.
\ea
\ee
It is found that
\be \label{8.9}
I(r_+) = I(r_-) =0.
\ee
When we expand $I(r)$ to the second order of $\frac{GM}{r}$,
it becomes a second order polynomial of $\frac{1}{r}$.
Two roots of $I(r)$ are $r_+$ and $r_-$, so $I(r)$ must
have the form
\be \label{8.10}
I(r) = c_1 \left ( \frac{1}{r_-} - \frac{1}{r} \right )
\left ( \frac{1}{r} - \frac{1}{r_+} \right ),
\ee
where $c_1$ is a constant. When $r$ approaches infinity,
$B(r) =1$. Combine (\ref{8.8}) and (\ref{8.10}), we get
\be \label{8.11}
c_1 = \frac{r_-^2 B^2(r_+) (B^2(r_-) -1)
- r_+^2 B^2(r_-) (B^2(r_+) -1)}
{r_+ r_- (B^2(r_+) - B^2(r_-) )}.
\ee
$B(r)$ is given by (\ref{4.30}), Using that result,
we find that $c_1$ have a simpler form
\be \label{8.12}
c_1 = 1- 2 G M \left (
\frac{1}{r_+} + \frac{1}{r_-}
\right ).
\ee
Inserting (\ref{8.10}) into (\ref{8.6}) gives
\be \label{8.13}
{\rm d}\varphi = c_1^{- \frac{1}{2}} A(r)
\left[
\left ( \frac{1}{r_-} - \frac{1}{r} \right )
\left ( \frac{1}{r} - \frac{1}{r_+} \right )
\right ]^{- \frac{1}{2}}
\frac{{\rm d}r}{r^2}.
\ee
$A(r)$ is given by (\ref{4.29}), in linear order approximation,
it is
\be \label{8.14}
 A(r) = \frac{1}{\sqrt{1-\frac{2 G M}{r}}}
 = 1 + \frac{G M}{r} + \cdots,
\ee
where $\cdots$ represents higher order terms. Inserting (\ref{8.14})
into (\ref{8.13}), we get
\be \label{8.15}
{\rm d}\varphi = - c_1^{- \frac{1}{2}}
\frac{\left ( 1 + \frac{G M}{r} \right )
{\rm d} \left ( \frac{1}{r} \right ) }{
\left\lbrack
\left ( \frac{1}{r_-} - \frac{1}{r} \right )
\left ( \frac{1}{r} - \frac{1}{r_+} \right )
\right \rbrack ^{\frac{1}{2}} }.
\ee
\\

Define
\be \label{8.16}
\frac{1}{r} \define \frac{1}{2}
\left ( \frac{1}{r_+} + \frac{1}{r_-} \right )
+  \frac{1}{2}
\left ( \frac{1}{r_+} - \frac{1}{r_-} \right )
\sin\psi .
\ee
At perihelia $\psi = - \frac{\pi}{2}$, while at aphelia
$\psi =  \frac{\pi}{2}$. From (\ref{8.16}), we have
\be \label{8.17}
\left ( \frac{1}{r_-} - \frac{1}{r} \right )
\left ( \frac{1}{r} - \frac{1}{r_+} \right )
= \frac{1}{4}
\left ( \frac{1}{r_-} - \frac{1}{r_+} \right )^2
\cos^2 \psi.
\ee
Then, (\ref{8.15}) is changed into
\be \label{8.18}
{\rm d}\varphi =  c_1^{- \frac{1}{2}}
\left \lbrack
1 + \frac{G M}{2}
\left ( \frac{1}{r_+} + \frac{1}{r_-} \right )
+ \frac{G M}{2}
\left ( \frac{1}{r_+} - \frac{1}{r_-} \right )
\sin \psi
\right \rbrack
{\rm d}\psi.
\ee
Integrate (\ref{8.18}) from perihelia $r_-$ to distance $r$, we get
\be \label{8.19}
\varphi (r) -  \varphi (r_-)
=  c_1^{- \frac{1}{2}}
\left \lbrack
(\psi + \frac{\pi}{2})+ \frac{G M}{2}
\left ( \frac{1}{r_+} + \frac{1}{r_-} \right )
(\psi + \frac{\pi}{2})
- \frac{G M}{2}
\left ( \frac{1}{r_+} - \frac{1}{r_-} \right )
\cos \psi
\right \rbrack.
\ee
So, we have
\be \label{8.20}
\varphi (r_+) -  \varphi (r_-)
=  c_1^{- \frac{1}{2}}
\left \lbrack
1+ \frac{G M}{2}
\left ( \frac{1}{r_+} + \frac{1}{r_-} \right )
\right \rbrack  \pi.
\ee
Inserting (\ref{8.12}) into above equation gives
\be \label{8.21}
\varphi (r_+) -  \varphi (r_-)
=  \left \lbrack
1+ \frac{3 G M}{2}
\left ( \frac{1}{r_+} + \frac{1}{r_-} \right )
\right \rbrack  \pi.
\ee
The precession per revolution is
\be \label{8.22}
\Delta \varphi =
2 (\varphi (r_+) -  \varphi (r_-)) - 2 \pi
=  \frac{6 \pi G M}{2}
\left ( \frac{1}{r_+} + \frac{1}{r_-} \right ) .
\ee
The semilatus rectum $L$ is defined by
\be \label{8.23}
\frac{1}{L} \define  \frac{1}{2}
\left ( \frac{1}{r_+} + \frac{1}{r_-} \right ).
\ee
Then $\Delta \varphi$ is simplified to
\be \label{8.24}
\Delta \varphi =   \frac{6 \pi G M}{L} .
\ee
This result is also the same as that in  general relativity.\\

\section{The Time Delay of Radar Echoes Passing the Sun }
\setcounter{equation}{0}

Using the following relations
\be \label{9.1}
 p^r = \frac{p^r}{p^t} p^{t}
 = p^{t} \frac{{\rm d} r}{{\rm d} t}
 = \frac{E_0}{B^2(r)} \frac{{\rm d} r}{{\rm d} t},
\ee
and (\ref{7.7}), we can change equation (\ref{6.28})  into
\be \label{9.2}
 \frac{ E_0^2 A^2(r)}{B^4(r)}  \left (
 \frac{{\rm d} r}{{\rm d} t} \right )^2
 + \frac{J^2 }{r^2} - \frac{E_0^2 }{ B^2(r)}  = 0.
\ee
\\

At distance $r_0$ of the closest approach to the sun,
$\frac{{\rm d}r}{{\rm d}t}$ vanishes, so equation
(\ref{9.2}) gives
\be \label{9.3}
 J^2  = \frac{E_0^2 r_0^2}{ B^2(r_0)} .
\ee
Inserting (\ref{9.3}) into (\ref{9.2}) gives
\be \label{9.4}
 \frac{ E_0^2 A^2(r)}{B^4(r)}  \left (
 \frac{{\rm d} r}{{\rm d} t} \right )^2
 + \frac{E_0^2 r_0^2 }{B^2(r_0) r^2} - \frac{E_0^2 }{ B^2(r)}  = 0.
\ee
In above equation, $E_0^2$ can be cancelled, so we get
\be \label{9.5}
 \frac{  A^2(r)}{B^4(r)}  \left (
 \frac{{\rm d} r}{{\rm d} t} \right )^2
 + \frac{ r_0^2 }{B^2(r_0) r^2} - \frac{1 }{ B^2(r)}  = 0.
\ee
By multiplying this equation with $B^2(r)$, we may write it as
\be \label{9.6}
 \frac{  A^2(r)}{B^2(r)}  \left (
 \frac{{\rm d} r}{{\rm d} t} \right )^2
 = 1 - \frac{B^2(r)}{B^2(r_0)}
 \left ( \frac{ r_0 }{ r} \right )^2 .
\ee
\\

$B(r)$ is given  by (\ref{4.30}). In first order approximation, we have
\be \label{9.7}
1 - \frac{B^2(r)}{B^2(r_0)}
       \left ( \frac{r_0}{r} \right )^2
= \left \lbrack 1- \left ( \frac{r_0}{r}
\right )^2  \right \rbrack
\left \lbrack 1 -
\frac{2 G M r_0}{r (r + r_0)} + \cdots
\right \rbrack,
\ee
where $\cdots$ represents higher order approximation.
Inserting (\ref{4.29}), (\ref{4.30}) and (\ref{9.7})
into (\ref{9.6}) and making first order approximation,
we  get
\be \label{9.8}
   \frac{{\rm d} t}{{\rm d} r}
 = \left \lbrack 1- \left ( \frac{r_0}{r}
\right )^2  \right \rbrack ^{- \frac{1}{2}}
\left ( 1 + \frac{2 G M}{r}
+ \frac{G M r_0}{r(r + r_0)}
\right ).
\ee
\\

Integrate (\ref{9.8}), we get the time required for light to go
from  from $r_0$ to $r$ or from $r$ to $r_0$
\be \label{9.9}
  t(r,r_0)
 = \int_{r_0}^{r} \left \lbrack 1- \left ( \frac{r_0}{r}
\right )^2  \right \rbrack ^{- \frac{1}{2}}
\left ( 1 + \frac{2 G M}{r}
+ \frac{G M r_0}{r(r + r_0)}
\right ) {\rm d}r .
\ee
Using the following integration formula
\be \label{9.10}
\int \left \lbrack 1- \left ( \frac{r_0}{r}
\right )^2  \right \rbrack ^{- \frac{1}{2}}
 {\rm d}r = \sqrt{r^2 - r_0^2},
\ee
\be \label{9.11}
\int \left \lbrack 1- \left ( \frac{r_0}{r}
\right )^2  \right \rbrack ^{- \frac{1}{2}}
 \frac{ {\rm d}r }{r}
 = {\rm ln} \left( r+ \sqrt{r^2 - r_0^2} \right),
\ee
and
\be \label{9.12}
\int \left \lbrack 1- \left ( \frac{r_0}{r}
\right )^2  \right \rbrack ^{- \frac{1}{2}}
 \frac{ r_0 {\rm d}r }{r(r+r_0)}
 = \sqrt{\frac{r-r_0}{r+r_0}},
\ee
we get
\be \label{9.13}
  t(r,r_0)
 = \sqrt{r^2 - r_0^2}
 + 2 G M {\rm ln} \frac{ r+ \sqrt{r^2 - r_0^2} }{r_0}
 + G M  \sqrt{\frac{r-r_0}{r+r_0}}.
\ee
\\

If light  travelled in straight lines at unit velocity,
the time required for light to go from $r_0$ to $r$ is
$\sqrt{r^2 - r_0^2}$. So, in this process, the excess time
delay is
\be \label{9.14}
\ba{rcl}
 \Delta t(r,r_0)& = & t(r,r_0) - \sqrt{r^2 - r_0^2} \\
& = & 2 G M {\rm ln} \frac{ r+ \sqrt{r^2 - r_0^2} }{r_0}
 + G M  \sqrt{\frac{r-r_0}{r+r_0}}.
 \ea
\ee
Suppose that a radar signal grazes the sun and is
reflected back to earth, the total excess time delay is
denoted by $(\Delta t)_{max}$. When radar signal just grazes
the sun, $r_0$ is about equal to the radius of the sun, that is
\be \label{9.15}
r_0 \simeq R_{\odot}.
\ee
The total excess time delay is
\be \label{9.16}
(\Delta t)_{max} = 2 [
\Delta t(r_{\oplus},R_{\odot})
+ \Delta t(r_{M},R_{\odot} ) ].
\ee
$R_{\odot}$ is much smaller than the distances $r_{\oplus}$
and $r_M$ of the earth and Mercury from the sun
\be \label{9.17}
\frac{ R_{\odot}}{r_{\oplus}} \simeq 0, ~~~
\frac{ R_{\odot}}{r_M} \simeq 0.
\ee
Under this approximation, we have
\be \label{9.18}
(\Delta t)_{max} = 4 G M \left (
1 + {\rm ln} \frac{4 r_{\oplus} r_M}{R^2_{\odot}}
\right ).
\ee
This result is the same as that in general relativity.
\\

\section{Summary and Discussions}

In this paper, classical tests of gauge theory of gravity are
discussed. All discussions are based field equation of gravitational
gauge field and Newton's second law of motion. Final
results are the same as those given by general relativity.
Quantitative results given by this paper is trivial, for
they are well-known in general relativity. But from the
discussions in this paper, we can obtain some important
qualitative conclusions on the nature of  gravity, that is,
gravity can be treated as a kind of physical interactions
in flat Minkowski space-time, and the equation of motion
of a mass point in gravitational field is given by Newton's
second law of motion. In this case, all kinds of fundamental
interactions can be described in the same way.
\\

In order to discuss classical motion of a particle, we need first
determine what is the equation of motion of a particle in
gravitational field. The basic logic in gauge theory of gravity
is that gravity is  a kind of fundamental
interactions in flat space-time. According to our knowledge on
classical mechanics, a logically natural conclusion is that
the equation of motion of a mass point in gravitational field
is given by Newton's second law of motion in a relativistic form.
Selecting geodesic equation as equation of motion of a test particle
is not logically consistent with basic spirit in gauge theory
of gravity. Therefore, though the field equation of gravitational
gauge field in gauge theory of gravity is the same as the Einstein's
field equation in general relativity, if the equation of motion of a
test particle in gauge theory of gravity is different from that in
general relativity, gauge theory of gravity will give different
results on classical tests of gravity. So, discussions on classical
tests of gravity in gauge theory of gravity are not a trivial task.
Fortunately, we found that, based on the classical solution of field
equation and Newton's second law of motion, gauge theory of gravity
gives out the same theoretical expectations on classical tests
of gravity as general relativity. An important physical conclusion
obtained from this work is that Newton's second law of motion
is also applicable to classical gravity. For a long time, we know that
Newton's second law of motion is not applicable to gravity,
especially from the point of view of general relativity. So,
in classical mechanics, Newton's second law of motion should be
a fundamental law which is applicable to all kinds of fundamental
interactions including gravity.
\\

After we written out Newton's second law of motion, the first task
in front of us is to determine the gravitational force acting
on the test particle. In this paper, a general relativistic
form of the gravitational force on a test particle is obtained,
which is the basis of  our discussions on classical problems
of gravity. As we have discussed in literature \cite{23}, gravitational
force contains Newtonian gravitational force which is transmitted
by gravitoelectric field and gravitational Lorentz force which
is transmitted by gravitomagnetic field. For a relativistic particle,
the gravitational force on it does not along the line connecting
the centers of mass of two bodies.
\\

In general relativity, the equation of motion of a mass point is given
by geodesic equation. In gauge theory of gravity, the equation of
motion of a mass point is given by Newton's second law of motion.
For spherical symmetric problems, these two equations are equivalent.
If the particle has inner spin, there is extra coupling between spin
and gravitomagnetic field\cite{12,13}, and there will extra gravitational
force in (\ref{5.13}) which comes from the coupling between spin
and gravitomagnetic field. Therefore, equation (\ref{5.13}), or
the geodesic equation in general relativity, is only applicable
to spinless particle. After consider spin of the test particle,
equation (\ref{5.13}) must be modified. A important consequence
of this modification is that the orbit of the motion dependents
on the spin of the particle, which violates weak equivalence
principle. This modification also affects the deflection of light
by the sun. But because the gravitomagnetic field generated by
the sun is too small, the influence from this modification
is too small to be detectable. \\

In gauge theory of gravity, space-time is always flat. In chapter 3, we
set up the formulation of gauge theory of gravity in curvilinear
coordinate system. In that case, space-time is still flat. The affine
connection defined is essentially different from that in general
relativity, for this affine connection does not contain effects
of gravity. \\

\end{document}